\documentclass[a4paper,11pt]{article}
\usepackage[utf8]{inputenc}
\usepackage[english]{babel}
\usepackage{amsmath}
\usepackage{amssymb}
\usepackage{jheppub}

\usepackage{physics}

\usepackage{amsmath,amssymb,amsfonts,mathrsfs}
\def\be#1\ee{\begin{align}#1\end{align}} 

\usepackage{hyperref}
\hypersetup{colorlinks=true,linkcolor=blue,citecolor=red,urlcolor=black,pdfauthor={Giulio Neri}}

\newcommand{\SISSA}{\affiliation{SISSA, International School for Advanced Studies, \\
via Bonomea 265, 34136 Trieste, Italy}}
\newcommand{\InfnTS}{\affiliation{INFN, Sezione di Trieste,\\
via Valerio 2, 34127 Trieste, Italy}}
\newcommand{\IFPU}{\affiliation{IFPU, Institute for Fundamental Physics of the Universe,\\
via Beirut 2, 34014 Trieste, Italy}}

\usepackage{comment}

\newcommand{\man}{\mathcal M}
\newcommand{\ren}{\text{ren}}
\newcommand{\rb}{\eta}
\newcommand{\pois}{\Sigma}

\newcommand{\prtl}{\text{bdry}}

\title{
On the resilience of the gravitational variational principle under renormalization
}

\author{Giulio Neri}
\author{and Stefano Liberati}

\SISSA\IFPU\InfnTS

\emailAdd{gneri@sissa.it}
\emailAdd{liberati@sissa.it}

\abstract{A well-defined variational principle for gravitational actions typically requires to cancel boundary terms produced by the variation of the bulk action with a suitable set of boundary counterterms. This can be achieved by carefully balancing the coefficients multiplying the bulk operators with those multiplying the boundary ones. A typical example of this construction is the Gibbons-Hawking-York boundary action that needs to be added to the Einstein-Hilbert one in order to have a well-defined metric variation for General Relativity with Dirichlet boundary conditions. Quantum fluctuations of matter fields lead to the renormalization of said coefficients which may or may not preserve this balance. Indeed, already at the level of General Relativity, the resilience of the matching between bulk and boundary constants is far from obvious and it is anyway incomplete given that matter generically induces quadratic curvature operators. We investigate here the resilience of the matching of higher-order couplings upon renormalization by a non-minimally coupled scalar field and show that a problem is present. Even though we do not completely solve the latter, we show that it can be greatly ameliorated by a wise splitting between dynamical and topological contributions. Doing so, we find that the bulk-boundary matching is preserved up to a universal term, whose nature and possible cancellation we shall discuss in the end.}

\arxivnumber{}

\begin{document}

\maketitle \flushbottom 

\section{Introduction}

There are a few aspects of modern theoretical physics that are as enigmatic and intriguing as black hole thermodynamics. When it emerged in the early seventies, it offered the first glance at a possible unification of quantum physics and classical General Relativity. Nonetheless, the thermodynamic behavior of gravitational objects such as black holes and the universal character of their entropy, which is independent of the number or type of matter fields interacting with them, represents an open question hinging at the very core of the relationship between gravity and matter. Nowadays, the most commonly accepted resolution of this so-called ``species problem" relies on the way the gravitational action is renormalized by quantum fluctuations of matter fields. To be precise, the species dependence is present in the entanglement entropy of quantum fields across the horizon, but it is hidden by the renormalization of the constants that enter the gravitational part of the entropy~\cite{Susskind:1994sm, Fursaev:1994ea, Frolov:1995xe, Frolov:1995pt, Larsen:1995ax, Frolov:1996aj, Frolov:1998vs,Cooperman:2013iqr}.

The entropy of black holes, and their thermodynamics in general, can be easily derived with a path integral formulation over geometries and fields that are periodic in (Euclidean) time~\cite{Gibbons:1976ue}.
In the context of a semiclassical approximation, one usually fixes a background geometry $\bar g_{\mu\nu}$ which is a solution of the classical theory of gravity (GR, for example), and then adds matter fields on top of that to study how quantum fluctuations backreact on the geometry. Gravitational perturbations can be treated as yet another matter field propagating on the background. In order for this to make sense, the background solution should correspond to a stationary point of the (bare) classical action $S_\text{cl}$\footnote{We will use $S_\text{cl}$ to refer to the classical action as a functional parametrized by some coefficients, and $S_0$ or $S_{\ren}$ to refer respectively to the same functional with bare or renormalized coefficients.}. Once matter fields are integrated out, the effect of their fluctuations produces an effective contribution to the gravitational action $W_m(\bar g)$ which typically contains divergences corresponding to the chosen background. Upon regularization, one can absorb the troublesome terms through the renormalization of fields and couplings. What is left behind is a finite action $S_\ren(\bar g)=S_0(\bar g)+\text{div}\{W_m(\bar g)\}$ that one can think of as a functional $S_\ren[g]$ evaluated on $g=\bar g$.
Unfortunately, there is no guarantee that the stationary points of this renormalized action, even neglecting quantum effects, are small departures from the chosen background $\bar g_{\mu\nu}$ expressed in terms of renormalized couplings. Indeed, as the following example proves, not even the simultaneous existence of those stationary points under the same boundary conditions is a trivial assumption. If we assume that all renormalized gravitational constants, except for Newton's constant, vanish, Minkowski spacetime is a stationary point for the renormalized action. However, unless some particular kind of symmetry is present, the bare constants do not vanish and the stationary points of the bare action may not even correspond to asymptotically flat spacetimes.

This issue is often overlooked in the literature. What is common is to assume that it is possible to work with a finite classical action by just replacing each bare coupling constant with its renormalized (observed) value. In this lore, renormalization has been made from the very beginning and the chosen background $\bar g_{\mu\nu}$ is taken to be a stationary point of the renormalized classical action.
This approach might appear reasonable since physical observables are oblivious to the renormalization procedure that we use to obtain our predictions, and sensible questions only refer to renormalized quantities. Accordingly, one might say that only the renormalized classical action, and not the bare one, needs to have a well-posed variational principle. However, we believe that there are good reasons for not being satisfied with this argument, one being that renormalization is not merely the absorption of artificial divergences into unobservable bare constants, but it also entails observable effects such as the flow of couplings with the energy scale.

The compatibility between the stationary points of the bare action and those of the renormalized one has been explicitly checked for manifolds without boundaries. On the other hand, the extension to manifolds with boundaries has yielded contrasting results so far, and, to the best of our knowledge, it has not yet been carried out completely. In the latter case, the root of the problem is that the bulk and boundary gravitational constants have different beta functions in principle. Due to this difference, any relationship that one can fix at a specific scale based on variational principle arguments is likely to be disrupted during the process of renormalization. Ideally, one would like to have a resilient matching that can hold at any scale.
This has been noted a while ago for both the cosmological and Newton's constant~\cite{Barvinsky:1995dp, Becker:2012js}, but, to the best of our knowledge, it has not found a definite solution yet. In~\cite{Becker:2012js}, it was claimed that there is no problem in having independent constants in the bulk and in the boundary because they enter different observables.
Albeit simple, such a resolution has been shown to be unreliable as the mismatch itself was based on non-diffeomorphism invariant boundary conditions and it has been amended in~\cite{Falls:2017cze}. 
More recently, the authors of~\cite{Jacobson:2013yqa} showed that one can find suitable boundary conditions for matter fields (scalars, gauge fields, and gravitons) so that the two Newton's constants (the bulk and the boundary one) are renormalized in the same way at one-loop in perturbation theory. It thus seems that a robust variational principle can be set up for almost all relevant matter fields. However, it has been pointed out (see footnote $11$ of Ref.~\cite{Jacobson:2013yqa}), it is crucial to check that whatever solution one finds keeps working when higher-order couplings are taken into account. This is because these couplings are expected to be generated by radiative corrections from matter in curved spacetimes.

Hence, in this paper, we study the renormalization of higher-order constants in the simple case of a non-minimally coupled scalar field and show that the idea found in~\cite{Jacobson:2013yqa} cannot be extended to them, at least in the standard setting of the boundary value problem for the metric tensor. 
Then, in order to move forward, we emphasize the relevance of properly identifying the actual dynamical degrees of freedom that should be fixed at the boundary. Even though we may not offer a definitive solution to the general problem, we gain valuable insights that we hope will serve as guidance in the resolution of this long-standing and often neglected problem.

The structure of the paper is the following. In Sec.~\ref{variational principle}, we introduce the theoretical foundations of our work, stating what we define to be a well-posed variational principle in a manifold with boundaries and introducing the concept of \textit{bulk-boundary matching}. We also review the heat kernel regularization and how renormalization might spoil said matching.
In Sec.~\ref{sec:GA}, we first repeat the analysis of Newton's constant renormalization and point out that the idea proposed in Ref.~\cite{Jacobson:2013yqa} for preserving the corresponding matching in the presence of a non-minimally coupled scalar field, faces a difficulty when one tries to implement the needed boundary conditions. We then investigate gravitational actions with higher-order couplings and in particular the case of quadratic gravity. By using an auxiliary fields method we obtain a boundary action that generalizes the standard Gibbons-Hawking-York boundary term of General Relativity. Considering the one-loop effective action of a scalar field, we show that there is no way to obtain a bulk-boundary matching that survives renormalization. We then look for a better solution. In Sec.~\ref{sec:improved}, starting from the example of Lanczos-Lovelock theories, we construct an improved boundary action that has the advantage of properly identifying the truly dynamical degrees of freedom. With this prescription, we end up with a bulk-boundary matching which is almost preserved at one loop. Indeed, what we obtain is that all the relevant matching conditions are preserved by renormalization except for one which concerns a universal boundary term that depends solely on the extrinsic curvature. Before moving to conclusions, we comment on this partial solution and provide an argument to get rid of the remaining term. Finally, in Sec.~\ref{sec:CN}, we summarize our findings and discuss potential concerns regarding the limitations of our work and its compatibility with previous results on the renormalization of Newton's constant.

Throughout the paper, we work in a four-dimensional Riemann spacetime with Euclidean signature. For the reader's convenience, the conventions we follow in this work, along with the definition of other familiar objects, have all been collected in Appendix~\ref{app:GI}.

\section{Preliminaries}
\label{variational principle}
In this section, we shall review the basic theoretical background for our investigation, starting with the formulation of the variational principle. It is well known that on a manifold $\man$ with a non-empty boundary $\partial\man$, a generic action can be written as the sum of a bulk term $S^{\text{bulk}}$ and a boundary one $S^\prtl$. Upon variation, one gets the following structure
\begin{equation}
    \delta S[\Psi]=\delta S^{\text{bulk}}[\Psi]+\delta S^{\prtl}[\Psi]=\int_\man \mathcal E[\Psi]\delta \Psi+\Theta[\Psi,\delta \Psi]+\delta S^\prtl[\Psi],
\end{equation}
where $\Psi$ denotes the collection of dynamical fields, $\mathcal E$ is the corresponding collection of field equations, and $\Theta\equiv\int_\man\dd\theta=\int_{\partial\man}\theta$ is the spacetime integral of the so-called symplectic potential, which is that part of the bulk action variation that can be written as a total derivative. A well-posed variational principle requires that stationary points of the action $\delta\mathcal S[\bar\Psi]=0$ correspond to the solutions of the field equations $\mathcal E[\bar\Psi]=0$. Equivalently, this means that the boundary contribution $\Theta+\delta S^\prtl$ must vanish on-shell provided that we impose proper boundary conditions.

\subsection{The bulk-boundary matching resilience problem}

When the above discussion is applied to the concrete case of General Relativity (GR), a well-posed variational problem with Dirichlet boundary conditions for the metric tensor requires that we supplement the Einstein-Hilbert (EH) action with the so-called Gibbons-Hawking-York (GHY) boundary term~\cite{Gibbons:1976ue}, which depends on the trace of the extrinsic curvature $K$, so that the total action reads
\begin{equation}
S^{\text{GR}}[g]=-\frac{1}{16\pi G}\int_{\cal M}R-\frac{2\sigma}{16\pi G} \int_{\partial \man} K,
\end{equation}
where $\sigma$ is the norm of the unit normal to the boundary which for us is always equal to $+1$ since we work in the Euclidean spacetime. For the sake of conciseness, we have avoided writing explicitly the volume forms when integrating over $\man$ or $\partial\man$. Note that, in calculations where the actual value of the action is needed, one might need to normalize the boundary term by subtracting the analogous term with $K\to K_0$ corresponding to the embedding of $\partial\man$ in a reference maximally symmetric version of the spacetime solution one is considering. Here, we shall omit it.

The Gibbons-Hawking-York term plays an important role in black hole thermodynamics as it is the only non-vanishing contribution to the free energy when the GR action is evaluated on-shell. The coefficient in front of this term must have a precise value if we want it to cancel the residual boundary terms coming from the bulk action variation (cfr. Sec.~\ref{sec:D2}). We refer to this tuning as \textit{bulk-boundary matching}.

If we add more terms to the gravitational action, as we are forced to do when taking into account the effect of matter fluctuations, the boundary action will likewise acquire new terms and require the imposition of new boundary conditions. This is usually interpreted as signaling the presence of new propagating degrees of freedom that one can try to make manifest with a different parametrization of the field content (see e.g.~\cite{Belenchia:2016bvb}). For example, it is known that $f(R)$ gravity is actually a scalar-tensor theory in disguise: the first sign of this is that we need to fix $f'(R)$ on the boundary to cancel all the boundary contributions from the bulk action variation~\cite{Sotiriou:2008rp, DeFelice:2010aj}. A second possible diagnosis is offered by Jacobson's derivation of the gravitational field equations~\cite{Jacobson:1995ab} from the Clausius relation in a local Rindler wedge. In that framework, new degrees of freedom manifests as new channels to dissipate excitations of the local horizon~\cite{Eling:2006aw, Chirco:2009dc}.

Of course, introducing higher-order operators will require new bulk-boundary matching conditions between the corresponding couplings. The problem we want to address is whether such matching is resilient against the aforementioned renormalization procedure, given that the latter modifies both bulk and boundary couplings of the theory. As we will show in the next section, the way in which boundary couplings are modified is sensitive to the boundary conditions that matter fields are subjected to, while the same is not true for bulk couplings. It is thus far from trivial that some matching can be preserved.

\subsection{Heat kernel and renormalization}
\label{sec:HK}

In this paper, we will limit our analysis to the quantum theory of a scalar field coupled to gravity. The partition function is formally given by the Euclidean path integral
\begin{equation}
\label{partition function}
    Z=\int[d g][d\phi]e^{-S_\text{cl}[g]+S_m[\phi;\,g]},
\end{equation}
where the integration is over field configurations periodic in Euclidean time and subject to suitable boundary conditions on the spacetime boundary $\partial\man$. In the previous formula, we split the classical action into a purely gravitational part $S_\text{cl}$ and a part $S_m$ describing the scalar field propagation and its interaction with gravity. The integral of the latter yields the so-called matter effective action
\begin{equation}
    W_m[g]\equiv -\ln\int[d\phi]e^{-S_m[\phi;\,g]},
    \label{eq:EffAct}
\end{equation}
so that the path integral reduces to
\begin{equation}
\label{gravity partition function}
    Z=\int[d g] e^{-S_{\text{cl}}[g]-W_m[g]}.
\end{equation}
If the action $S_m$ is quadratic in the field $\phi$ (or we consider only small perturbations around a classical solution), we can perform the integral \eqref{eq:EffAct} explicitly and get the formal result
\begin{equation}
\label{eq. matter eff action}
    W_m[g]=\frac{1}{2}\ln\det\Delta,
\end{equation}
where $\Delta=\fdv[2]{S_m}{\phi}\eval_{\phi=0}$ ($\phi=0$ is the classical solution we are considering). In this work, we take this operator to have the Laplace form appropriate for a generically-coupled scalar field
\begin{equation}
    \label{matter eom}
    \Delta=-\nabla^2+\xi R,
\end{equation}
with $\nabla_\mu$ being the covariant derivative compatible with $g_{\mu\nu}$. In the presence of a submanifold like $\partial\man$, there is another notion of covariant derivative $\mathcal D_a$ which is the one compatible with the induced metric $h_{ab}$. Furthermore, we assume that the scalar field satisfies one of the following boundary conditions
\begin{subequations}
    \begin{gather}
    \label{Dirichlet bc}
    \phi\,|_{\partial\man}=0,\quad\text{Dirichlet b.c.}\\
    (\nabla_n+\rb K)\phi\,|_{\partial\man}=0,\quad\text{Generalized Neumann (or Robin) b.c.}
    \label{Robin bc}
    \end{gather}
\end{subequations}
where $\nabla_{n}=n^\mu \nabla_\mu$, $n^\mu$ is the outward normal vector to $\partial\man$, and $K$ is the trace of the extrinsic curvature. We can see that the parameter $\eta$ generalizes the pure Neumann boundary condition ($\nabla_n \phi|_{\partial\man}=0$) by allowing a dependence on the extrinsic curvature\footnote{The $\rb$ parameter is written as $-\xi/\alpha$ in~\cite{Jacobson:2013yqa}.}.

Written as in~\eqref{eq. matter eff action}, the effective action $W_m$ is ill-defined and needs to be regularized. A common procedure is to use the heat kernel regularization, which is based on the following integral representation, valid for any $\lambda>0$
\begin{equation}
    \lambda^{-s}=\frac{1}{\Gamma(s)}\int_0^\infty \frac{\dd t}{t^{1-s}}e^{-t\lambda},
\end{equation}
where $\Gamma(s)$ is the Gamma function. If we assume that all the eigenvalues of $\Delta$ are positive, we can expand its determinant as the product of these eigenvalues, write each logarithm as the derivative of $\lambda^{-s}$ at $s=0$, and apply the previous formula to get
\begin{equation}
    \ln\det\Delta=-\dv[]{}{s}\frac{1}{\Gamma(s)}\int_0^\infty \frac{\dd t}{t^{1-s}}\sum_\lambda e^{-t\lambda}\eval_{s=0}.
    \label{zeta func reg}
\end{equation}
As $s\to 0$, the integral diverges for small values of $t$\footnote{Small $t$ divergences correspond to UV divergences. For the sake of conciseness, we are neglecting possible divergences in the $t\to\infty$ limit.}. This suggests regularizing the expression by introducing a small distance cut-off $\epsilon^2$ in the lower limit of integration. Now that the integral is finite, we can exchange the derivative and the integral. Considering that $\Gamma^{-1}(s)\to 0$ and $(\Gamma^{-1}(s))'\to 1$ for $s\to 0$, we just drop the pre-factor and set $s=0$:
\begin{equation}
    \ln\det\Delta=-\int_{\epsilon^2}^\infty \frac{\dd t}{t}\sum_\lambda e^{-t\lambda}.
\end{equation}
We recognize the sum to be the trace of the heat kernel of the differential operator $\Delta$\footnote{The heat kernel is defined as the solution to the following Cauchy problem
\begin{equation}
\begin{cases}
    (\partial_t+\Delta_x)K_\Delta(x,y;t)=0,\\
    K_\Delta(x,y;0)=\delta(x,y),
\end{cases}
\end{equation}
supplemented by proper boundary conditions on $\partial\man$. Its trace $\Tr K_\Delta$ is obtained by taking the trace of the coincidence value $K_\Delta(x,x;t)$ over the vector bundle on which $\Delta$ acts and integrating it over spacetime.} so that the regulated effective action can be expressed as
\begin{equation}
    W_m[g]=-\frac{1}{2}\int_{\epsilon^2}^\infty \frac{\dd t}{t} \Tr K_\Delta (t).
    \label{eq:TrHK}
\end{equation}
Since we are interested in the renormalization of UV divergences, we can focus on the small $t$ regime and expand the heat kernel trace. It has been shown that, for $t\to 0$
\begin{equation}
    \Tr K_\Delta(t)=\frac{1}{(4\pi t)^2}\sum_{k=0}^\infty \mathfrak{a}_k t^{k/2}
\end{equation}
up to exponentially suppressed terms \cite{Vassilevich:2003xt}. Under the assumed boundary conditions for the scalar field, the coefficients $\mathfrak{a}_k$ can be written in terms of (integrals of) local geometrical invariants. For even $k$, they involve integration over the entire manifold, including $\partial\man$, while, for odd $k$, there is no bulk contribution.
Adapting from~\cite{Vassilevich:2003xt}, we rewrite the first few coefficients for Dirichlet (D) and Robin (R) boundary coefficients as
\begin{subequations}
\begin{align}
\label{aD0 coefficients}
    \mathfrak{a}^D_0=&\int_\man 1,\\
\label{aD1 coefficients}
    \mathfrak{a}^D_1=&-\frac{(4\pi)^{1/2}}{4}\int_{\partial\man} 1,\\
\label{aD2 coefficients}    
    \mathfrak{a}^D_2=&-\int_\man\pqty{\xi-\xi_c}R+\frac{1}{3}\int_{\partial\man}K,\\
\label{aD3 coefficients}
    \mathfrak{a}^D_3=&\,\frac{(4\pi)^{1/2}}{4}\int_{\partial\man}\pqty{(\xi-\xi_c)R-\frac{1}{12} R_{\mu\nu}n^\mu n^\nu+\frac{5}{48}G_2-\frac{11}{288}K^2},\\
\label{aD4 coefficients}
    \mathfrak a^D_4=&\,\frac{1}{120}\int_{\man} C_{\mu\nu\rho\sigma}C^{\mu\nu\rho\sigma}-\frac{4\pi^2}{45}\chi[\man]+\int_{\partial\man}\pqty{\frac{2}{35}G_3-\frac{1}{15}C_{\mu\nu\rho\sigma}n^\nu n^\sigma K^{\mu\rho}}+\nonumber\\
    &+\pqty{\xi-\xi_c}^2\int_{\man}\frac{R^2}{2}-\pqty{\xi-\xi_c}\int_{\partial\man}\pqty{\frac{1}{3} RK+\frac{1}{2}\nabla_n R},
\end{align}
\end{subequations}
and
\begin{subequations}
\begin{align}
\label{aR0 coefficients}
    \mathfrak{a}^R_0=&\int_\man 1,\\
\label{aR1 coefficients}
    \mathfrak{a}^R_1=&\,\frac{(4\pi)^{1/2}}{4}\int_{\partial\man} 1,\\
\label{aR2 coefficients}    
    \mathfrak{a}^R_2=&-\int_\man\pqty{\xi-\xi_c}R-\int_{\partial\man}\pqty{2\rb-\rb_c}K,\\
\label{aR3 coefficients}
    \mathfrak{a}^R_3=&-\frac{(4\pi)^{1/2}}{4}\int_{\partial\man}\biggr((\xi-\xi_c)R-\frac{1}{12} R_{\mu\nu}n^\mu n^\nu-\frac{1}{48}G_2-\frac{1}{32}K^2-(\eta-\eta_c)(2\eta-\eta_c)K^2\biggl),\\
\label{aR4 coefficients}
    \mathfrak a^R_4=&\,\frac{1}{120}\int_{\man} C_{\mu\nu\rho\sigma}C^{\mu\nu\rho\sigma}-\frac{4\pi^2}{45}\chi[\man]+\int_{\partial\man}\pqty{\frac{2}{45}G_3-\frac{1}{15}C_{\mu\nu\rho\sigma}n^\nu n^\sigma K^{\mu\rho}}+\nonumber\\
    &+\pqty{\xi-\xi_c}^2\int_{\man}\frac{R^2}{2}+\pqty{\xi-\xi_c}\int_{\partial\man}\pqty{\pqty{2\eta-\eta_c} RK+\frac{1}{2}\nabla_n R}+\nonumber\\
    &-\pqty{\eta-\eta_c}\int_{\partial\man}\pqty{\frac{2}{15}K G_2+\frac{4}{3}\pqty{\eta-\eta_c}^2 K^3}.
\end{align}
\end{subequations}
For later convenience, we wrote the geometrical invariants in a basis that makes manifest the properties under Weyl scaling $g_{\mu\nu}(x)\to \Omega^2(x) g_{\mu\nu}(x)$. Therefore we expressed the extrinsic curvature contractions in terms of the (first two) polynomials $G_n$ introduced by Melmed in \cite{Melmed_1988}
\begin{equation}
    G_2=K^a_b K^b_a-\frac{1}{3}K^2,\qquad G_3=K^a_b K^b_c K^c_a - K K^a_b K^b_a+\frac{2}{9}K^3,
    \label{eq:Gi}
\end{equation}
because they transform covariantly under such transformations\footnote{\label{Melmed poly} In general, one can construct $G_n\equiv\tr({}_\mathbb T K^n)$, where ${}_\mathbb T K_{ab}$ denotes the traceless part of $K_{ab}$. Under a Weyl scaling $g_{\mu\nu}(x)\to\Omega^2(x) g_{\mu\nu}(x)$, these polynomials transform covariantly as
\begin{equation}
    G_n(x)\to \Omega^{-n}(x) G_n(x).
\end{equation}
Given that the induced volume form scales as $\Omega^{d-1}$, the integral of $G_{d-1}$ is invariant (for us, $d=4$).
}. Likewise, we chose to write the Riemann tensor and its squared contractions in terms of the Weyl tensor $C_{\mu\nu\rho\sigma}$ and the Euler characteristic $\chi[\man]$. Finally, we used the special values $\xi_c=1/6$ and $\rb_c=1/3$ as a reference for the $\xi$- and $\eta$-dependent terms. These values are called conformal\footnote{For a generic dimension $d$, the conformal couplings are
\begin{equation}
    \xi_c(d)=\frac{d-2}{4(d-1)},\qquad \eta_c(d)=\frac{d-2}{2(d-1)}=2\xi_c(d).
\end{equation}} because they respectively make the equation of motion~\eqref{matter eom} and the Robin boundary condition~\eqref{Robin bc} invariant under Weyl scaling~\cite{Kennedy:1979ar}. Note that the Dirichlet boundary condition is trivially Weyl invariant.

Integrating Eq.~\eqref{eq:TrHK} over $t$, we obtain a similar expansion for the matter effective action
\begin{equation}
    W_m[g]=-\frac{1}{(4\pi)^{2}}\sum_{k=0}^\infty \frac{\mathfrak{a}_k}{(4-k)\epsilon^{4-k}}.
    \label{eq:IntW}
\end{equation}
This expression allows us to have better control over the divergences, which we recognize to correspond to the heat kernel coefficients with $k\le 4$. We remark that, even if we wrote each term in the same power-like form, the one with $k=4$ actually diverges logarithmically. The standard way to deal with these would-be infinities is to assume that the bare couplings in the classical action are formally infinite too so that divergences in matching operators in $S_0$ and $W_m$ cancel out against each other to yield a finite quantity. After that, the regulator can be dropped ($\epsilon\to 0$) and what remains in the exponent of Eq.~\eqref{gravity partition function} is the renormalized action $S_\ren=S_0+\text{div}\{W_m\}$.

As anticipated, divergences appear in front of both bulk and boundary operators with relative weights that have nothing to do ---~as far as we know~--- with the well-posedness of a variational principle. Therefore it is not obvious that whatever matching one imposes for the bare action persists after renormalization. In order to see explicitly how this can happen, we first consider the simple case of GR.

\section{Gravitational action and boundary term}
\label{sec:GA}

It is well known that GR is not renormalizable at one-loop (except for the case of pure gravity with no cosmological constant~\cite{tHooft:1974toh}) so that the bare GR action has to be supplemented with higher order operators. First, we assume that the bare couplings in front of these operators are tuned to cancel exactly the corresponding divergences from the matter effective action and consider only the renormalization of the cosmological constant and Newton's constant. Later, we will not ask for this tuning and shift our focus to higher-order coupling.

\subsection{Renormalization of dimension $\le 2$ operators}
\label{sec:D2}

As already discussed in the introduction, a well-defined metric variational principle for the EH action requires the addition of a GHY boundary term. Let us then write the bare gravitational action 
in the general form
\begin{equation}
    S^{GR}_{0}[g]=-\frac{1}{16\pi G_{0}}\int_\man (R-2\Lambda_0)-\frac{1}{8\pi \mathcal G_{0}}\int_{\partial\man}(K-2\lambda_{0})\,,
\end{equation}
where we allowed for possibly different bare gravitational constants in the bulk ($G_0,\Lambda_0$) and in the boundary ($\mathcal G_0,\lambda_0$). For generic, unrelated values of the constants, this action does not lead to a well-defined variational principle. If we vary the action and evaluate it on the solution of the vacuum equations of motion ($G_{\mu\nu}+\Lambda_{0} g_{\mu\nu}=0$), we obtain the following boundary contribution 
\begin{equation}
    \delta S^{GR}_{0}[g]=-\frac{1}{16\pi}\int_{\partial\man} \pqty{\frac{2}{\mathcal G_{0}}\delta K-\frac{1}{G_{0}} n_\mu \delta v^\mu+\frac{1}{\mathcal G_{0}}(K-2\lambda_{0}) h^{ab}\delta h_{ab}},
\end{equation}
with $\delta v^\mu$ a vector defined by $g^{\mu\nu}\delta R_{\mu\nu}=-\nabla_\mu \delta v^\mu$. With some extra steps which can be found in several textbooks, see e.g.~\cite{Wald:1984rg}, one can show that the term $n_\mu \delta v^\mu$ can be expanded as $2\delta K+K^{ab}\delta h_{ab}$, so that
\begin{equation}
    \delta S^{GR}_{0}[g]=-\frac{1}{16\pi}\int_{\partial\man} \pqty{2\delta K\pqty{\frac{1}{\mathcal G_{0}}-\frac{1}{G_{0}}}-\frac{1}{G_{0}} K^{ab}\delta h_{ab}+\frac{1}{\mathcal G_{0}}(K-2\lambda_0)h^{ab}\delta h_{ab}}.
\end{equation}
The usual way in which this residual contribution is set to zero is by imposing Dirichlet boundary conditions for the metric field $\delta h_{ab}=0$ and requiring the bulk and boundary Newton's constants to be the same, i.e. $\mathcal G_{0}=G_0$. This is the standard bulk-boundary matching in GR (see however e.g.~\cite{Oliveri:2019gvm} for an alternative matching procedure involving non-Dirichlet boundary conditions).

Let us now consider what happens when the contribution from the matter effective action is taken into account.
Following the procedure described in Sec.~\ref{sec:HK}, we see that, once matter is integrated out, the bulk cosmological constant absorbs the divergence multiplying the identity in the heat kernel coefficient $\mathfrak a_0$. Similarly, the boundary cosmological constant absorbs the coefficient of the identity in $\mathfrak a_1$. On the other hand, both Newton's constants are renormalized according to $\mathfrak a_2$
\begin{subequations}
\begin{align}
    \frac{\Lambda_\ren }{8\pi G_\ren }&=\frac{\Lambda_{0}}{8\pi G_{0}}-\frac{1}{64\pi^{2}\epsilon^4}, \label{eq:ReLamb}\\
    \frac{\lambda_\ren }{4\pi \mathcal G_\ren }&=\frac{\lambda_{0}}{4\pi \mathcal G_{0}}\pm\frac{1}{12(4\pi)^{3/2}\epsilon^3},\label{eq:Relamb}\\
    \frac{1}{16\pi  G_\ren }&=\frac{1}{16\pi  G_{0}}-\frac{1}{32\pi^2\epsilon^2}\pqty{\xi-\frac{1}{6}},\label{eq:ReG}\\
    \frac{1}{8\pi \mathcal G_\ren }&=\frac{1}{8\pi \mathcal G_{0}}-\frac{1}{32\pi^2\epsilon^2}\pqty{2\rb-\frac{1}{3}} \label{eq:RebG},
\end{align}
\end{subequations}
where the upper/lower sign is for the Dirichlet/Robin boundary condition and $\eta$ is to be dropped in the Dirichlet case.

We start by analyzing the consequences of this renormalization in the simple case of a minimally coupled scalar field, i.e. $\xi=0$. We can easily see that if we impose either the Dirichlet or the pure Neumann boundary condition on that field ($\rb$ vanishes in both cases), the matching $\mathcal G=G$, corresponding to Dirichlet boundary conditions for the metric, is preserved
\begin{equation}
    \frac{1}{G_{\ren}}=\frac{1}{G_0}+\frac{1}{12\pi\epsilon^2}\overset{G_0=\mathcal G_0}{=}\frac{1}{\mathcal G_0}+\frac{1}{12\pi\epsilon^2}=\frac{1}{\mathcal G_{\ren}}
\end{equation}
While promising, this satisfactory result is jeopardized once we assume a non-minimal coupling for the scalar field. Indeed, for $\xi\ne 0$, the standard matching $\mathcal G_0=G_0$ is not preserved by the renormalization if the scalar field satisfies the Dirichlet or the pure Neumann boundary condition as $\eta=0$ in both cases. 
One can observe that, if the matching had the form $\mathcal G_0=G_0(1-6\xi)$, it would have been preserved. However, such matching would imply a set of boundary conditions for the metric that, if existing, are different for different values of the coupling $\xi$\,. While we cannot rule them out, it is not clear what guiding principle could lead to this scenario.

Alternatively, instead of looking for different ($\xi$-dependent) boundary conditions for the metric, one can take advantage of the possibility to use the Robin (generalized Neumann) boundary condition for the scalar field, i.e.~$\rb\neq 0$. Given a generic non-minimally coupled scalar field, we simply have to impose that it satisfies the Robin boundary condition with $\rb=\xi$. In this case, the quantum correction in Eq.~\eqref{eq:RebG} is exactly twice that in Eq.~\eqref{eq:ReG}, which makes them have the same ratio of the bare terms they renormalize. This is indeed the ingenious proposal advanced in~\cite{Jacobson:2013yqa}. What one needs to check now is whether this solution can be extended beyond GR to construct resilient bulk-boundary matching for higher-order couplings.
However, as we show in Appendix~\ref{app:scalar}, 
there is a difficulty in enforcing $\eta=\xi$ in a framework where both the metric and the scalar field are varied simultaneously. Indeed, when considering the action for the scalar field, one is only allowed to mix Neumann and Dirichlet boundary conditions with a constant coefficient, whereas the condition $\eta=\xi$ requires such coefficient to be proportional to the ({\em a priori} non-constant) mean extrinsic curvature $K$. 
In what follows, given that we cannot exclude that such difficulty can be overcome, for example by adding higher-order boundary terms, we shall nonetheless entail this possibility.

\subsection{Renormalization of dimension $4$ operators}
\label{sec:D4}

As we mentioned at the beginning of this section, a theory like GR, with operators of mass dimension $\le 2$ only, is not renormalizable because dimension $4$ operators are generated at one-loop level. Hence, we cannot content ourselves with the analysis done so far and we necessarily have to study the renormalization of the coefficients of such higher-order operators too. Hence, we now focus on the quadratic operators which supplement GR. 

We start by noticing that when we deal with operators that are quadratic in the bulk curvature, we also have to add operators of dimension $3$ on the boundary. In a metric theory without torsion or non-metricity, the bulk operators we have to consider are just three, namely
\begin{equation}
    \label{dim4 bulk operators}
    R^2,\quad R_{\mu\nu}R^{\mu\nu},\quad R_{\mu\nu\rho\sigma}R^{\mu\nu\rho\sigma},
\end{equation}
while the possible boundary operators are ten
\begin{equation}
    \label{dim3 bdry operators}
    \begin{split}
    &K^3,\quad K K^a_b K^b_a,\quad K^a_b K^b_c K^c_a,\quad R K, \quad R_{ab}K^{ab}, \quad R_{\mu\nu}n^\mu n^\nu K,\\
    &\quad R_{\mu\nu\rho\sigma}n^\mu n^\sigma K^{\mu\rho},\quad \nabla_n R,\quad n_\mu n_\nu \nabla_n R^{\mu\nu},\quad n_\mu h^\lambda_\nu \nabla_\lambda R^{\mu\nu}.
    \end{split}
\end{equation}
In this counting, we integrated every total derivative, which means that operators such as $\nabla^2 R$ are viewed as boundary operators, whereas operators such as $\mathcal D^2 K$ are simply discarded (the boundary $\partial\man$ has no boundary). Lastly, we can use the twice-contracted Bianchi identity to show that
\begin{equation}
    n_\mu n_\nu \nabla_n R^{\mu\nu}-\frac{1}{2}\nabla_n R=-n_\mu h^\lambda_\nu \nabla_\lambda R^{\mu\nu}=R_{\mu\nu}K^{\mu\nu}-R_{\mu\nu}n^\mu n^\nu K
\end{equation}
up to a total boundary derivative that we discard. Hence, without loss of generality, we can consider only three bulk operators and eight boundary ones. Instead of using~\eqref{dim4 bulk operators} and~\eqref{dim3 bdry operators} though, it is convenient to express the quadratic part of the action in the same basis that we used for~\eqref{aD4 coefficients} and~\eqref{aR4 coefficients} so that
\begin{equation}
\begin{split}
\label{QG action bulk}
    S_\text{cl}^{\text{bulk}}[g]=\int_\man \pqty{\alpha R^2+\beta C^2+\gamma E_4},
\end{split}
\end{equation}
and for what concerns the boundary action
\begin{equation}
\begin{split}
\label{QG action boundary}
 S_\text{cl}^{\prtl}[g]=\int_{\partial\man}\pqty{ {}^1{a}\, G_3+{}^2 a\, G_2 K+{}^3 a\, K^3 + {}^1 b\, B_3+{}^2 b\, \mathcal R K+{}^1c\, U+{}^2 c\, R K+{}^3c\, \nabla_n R}.
\end{split}
\end{equation}

Let us spell out in detail the above formulas in its notation and our choices in grouping its terms.
In the bulk, we chose to organize the three invariants in the so-called Weyl basis, where we defined
\begin{equation}
    C^2\equiv C_{\mu\nu\rho\sigma}C^{\mu\nu\rho\sigma}\,,\qquad E_4\equiv R^2-4R_{\mu\nu}R^{\mu\nu}+R_{\mu\nu\rho\sigma}R^{\mu\nu\rho\sigma}\,,
\end{equation}
respectively the squared Weyl tensor and the Gauss-Bonnet term (the name $E_4$ comes from the fact that this is also the Euler density in $4$d). In the boundary, we used different names for the coefficients in order to distinguish the three following subsets of operators
\begin{description}
    \item[$(a)$] Operators that are cubic in the extrinsic curvature. We again used the polynomials introduced by Melmed in~\cite{Melmed_1988}.
    \item[$(b)$] Operators that can be built out of the boundary intrinsic and extrinsic curvatures. If we denote by $\mathcal R_{ab}$ the Ricci curvature of $\partial\man$, we have $\mathcal R\equiv h^{ab}\mathcal R_{ab}$ and the following expression for the Chern-Simons form
\begin{equation}
    B_3=8\det K+8\pqty{\mathcal R_{ab}-\frac{1}{2}\mathcal R h_{ab}} K^{ab}, \quad \det K=\frac{1}{6}K^3-\frac{1}{2} K K^a_b K^b_a+\frac{1}{3}K^a_b K^b_c K^c_a.
    \label{eq:Chern}
\end{equation}
    \item[$(c)$] Operators that are written in terms of the full bulk curvature. We used the short-hand
\begin{equation}
    U\equiv C_{\mu\nu\rho\sigma}n^\mu n^\rho K^{\nu\sigma}.
\end{equation}
\end{description}

As soon as we add terms that contain two or more powers of the curvature to the usual GR action, the resulting equations of motion for the metric will be generically fourth-order\footnote{An exception being the $\alpha=\beta=0$ case, which is called Gauss-Bonnet gravity~\cite{Padmanabhan:2013xyr}; we will come back to it in Sec.~\ref{sec:LL}.}, hence propagating more degrees of freedom (d.o.f.) than just the two metric ones. These additional d.o.f. will be subjected to some boundary conditions in order to select the physically relevant solutions. In turn, these conditions will reflect on particular relations between the eight coefficients in \eqref{QG action boundary} and the three in \eqref{QG action bulk}. In the absence of these, the variational principle will not be able to select the solutions of the equations of motion.
Indeed, it is well known that the addition of the quadratic terms \eqref{dim4 bulk operators} generically implies six extra d.o.f. that, when expanded around $4$d flat spacetime, are associated with a massive traceless-tensor/graviton (five d.o.f) that is also a ghost~\cite{Stelle:1977ry}, and with a massive scalar (one d.o.f.), see e.g.~\cite{Deruelle:2009zk}. See also~\cite{Belenchia:2016bvb, Salvio:2018crh, Donoghue:2021cza} for extensive reviews and comments about the problem of ghosts generically associated with this class of theories.

While nothing prevents us from solving the fourth-order field equations once we are provided with some boundary conditions, it is preferable to have an understanding of the solutions in terms of phase space variables. Moreover, boundary conditions are meant to describe the physical processes that occur when the d.o.f. of the theory interact with the boundary. Therefore, it is more natural to first identify the present d.o.f. and then assign boundary conditions to them.
For these reasons, we present an alternative formulation of the theory based on the introduction of auxiliary fields which will help us identify the new degrees of freedom and their boundary conditions.

\subsection{Boundary action by auxiliary fields method}

Following the approach adopted in \cite{Deruelle:2009zk,Teimouri:2016ulk}, we introduce an auxiliary field $\varrho$ and a Lagrange multiplier $\Phi$, both with the same symmetries as the Riemann tensor, and rewrite the bulk action~\eqref{QG action bulk} as
\begin{equation}
    S^{'\text{bulk}}_\text{cl}[g,\varrho,\Phi]=\frac{1}{2}\int_{\man}\pqty{f(\varrho_{\mu\nu\rho\sigma})+\Phi^{\mu\nu\rho\sigma}(R_{\mu\nu\rho\sigma}-\varrho_{\mu\nu\rho\sigma})}.
\end{equation}
Also in this case we need to complement this action with a boundary term whose form we will determine.

Let us start by noticing that, thanks to the introduction of auxiliary fields which are independent of the metric, this action contains now second derivatives only at the linear level, e.g.~through the Riemann tensor. It is then easy to compute the variation and identify the symplectic potential, integrating by parts the Riemann tensor variation twice
\begin{equation}
\begin{split}
    \delta S^{'\text{bulk}}_{\text{cl}}[g,\varrho,\Phi]=\frac{1}{2}\int_\man\bigl[&\,\mathcal E_{\mu\nu}\delta g^{\mu\nu}+\pqty{\fdv{f}{\varrho_{\mu\nu\rho\sigma}}-\Phi^{\mu\nu\rho\sigma}}\delta \varrho_{\mu\nu\rho\sigma}+(R_{\mu\nu\rho\sigma}-\varrho_{\mu\nu\rho\sigma})\delta\Phi^{\mu\nu\rho\sigma}\bigr]+\\
    &+\int_{\partial\man}n_\sigma(\Phi^{\mu\nu\rho\sigma}\nabla_\nu\delta g_{\mu\rho}-\nabla_\nu \Phi^{\mu\nu\rho\sigma}\delta g_{\mu\rho}),
    \label{eq:varauxact}
\end{split}
\end{equation}
where $\mathcal E_{\mu\nu}$ are the field equations for the metric and the extra d.o.f. which only contains second derivative of $\Phi$. 
On the other hand, the equations of motion for the auxiliary fields are algebraic, so we could easily solve for them
\begin{equation}
    \varrho_{\mu\nu\rho\sigma}=R_{\mu\nu\rho\sigma},\qquad \Phi^{\mu\nu\rho\sigma}=\fdv{f}{\varrho_{\mu\nu\rho\sigma}}\eval_{\varrho=R},
\end{equation}
and replace them in $\mathcal E_{\mu\nu}$ to get back the original fourth-order equation for the metric. It is obvious that replacing $\varrho$ with its solution at the level of the action gives back the original one. What is less obvious is that one can replace the Lagrange multiplier $\Phi$ with its on-shell value in the action and obtain an equivalent one that is a functional of $g$ an $\varrho$ only~\cite{Faddeev:1988qp}. This can be done as long as the Hessian $\delta^2 f/\left( \delta \varrho_{\alpha\beta\gamma\delta}\delta \varrho_{\mu\nu\rho\sigma}\right)$ is non-singular. In the following, we assume this is the case and use $\Phi$ either as an independent field or as a shorthand for the derivative of $f$ quite interchangeably.\footnote{If the action we consider is quadratic in the curvature as in~\eqref{QG action bulk}, the Hessian of $f$ is regular as long as $\gamma\ne 0$, $\beta+\gamma\ne 0$, and $6\alpha+\gamma\ne0$. To see how one can obtain these conditions, let us go to the so-called Ricci basis~\cite{SingerThorpe+1970+355+366} of curvature invariants: $f=o_1 S^2+o_2 E^2+o_3 C^2$, with $o_1=6\alpha+\gamma$, $o_2=-\gamma$, $o_3=\beta+\gamma$. These terms are the squared contractions of
\begin{align}
    S_{\mu\nu\rho\sigma}&=\frac{R}{12}(g_{\mu\rho}g_{\nu\sigma}-g_{\mu\sigma}g_{\nu\rho}),\\
    E_{\mu\nu\rho\sigma}&=\frac{1}{2}(g_{\mu\rho} R_{\nu\sigma}-g_{\mu\sigma}R_{\nu\rho}+g_{\nu\sigma}R_{\mu\rho}-g_{\nu\rho}R_{\mu\sigma})+\frac{R}{12}(g_{\mu\rho}g_{\nu\sigma}-g_{\mu\sigma}g_{\nu\rho}),\\
    C_{\mu\nu\rho\sigma}&=R_{\mu\nu\rho\sigma}-\frac{1}{2}(g_{\mu\rho} R_{\nu\sigma}-g_{\mu\sigma}R_{\nu\rho}+g_{\nu\sigma}R_{\mu\rho}-g_{\nu\rho}R_{\mu\sigma})-\frac{R}{6}(g_{\mu\rho}g_{\nu\sigma}-g_{\mu\sigma}g_{\nu\rho}).
\end{align}
Since each of these tensors is orthogonal to the others, the Hessian of $f$ is diagonal in this basis. Hence its determinant is given by a product of the coefficients $o_i$ (each raised to some positive power) and will be non-zero as long as none of the $o_i$ vanish.}

Let us now focus on the integrated symplectic potential that can be read from the second line of Eq.~\eqref{eq:varauxact}
\begin{equation}
\Theta_g\equiv\int_{\partial\man}n_\sigma\left(\Phi^{\mu\nu\rho\sigma}\nabla_\nu\delta g_{\mu\rho}-\nabla_\nu \Phi^{\mu\nu\rho\sigma}\delta g_{\mu\rho}\right).
\label{eq:simPot}
\end{equation}
If we project the indices of $\Phi$ either normally or tangentially to $\partial\man$, we can define the following tangential tensors\footnote{The reader should not confuse these projections with those of e.g.~\cite{Deruelle:2009zk}, where tensors are projected normally and tangentially to a constant time hypersurface in an ADM decomposition of spacetime.}
\begin{align}
    \phi^{abcd}\equiv&\, \Phi^{\alpha\beta\gamma\delta}\,{e^a}_\alpha \,{e^b}_\beta \, {e^c}_\gamma \, {e^d}_\delta,\\
    \varphi^{abc}\equiv&\, \Phi^{\alpha\beta\gamma\delta} \,{e^a}_\alpha \,{e^b}_\beta \,{e^c}_\gamma \,n_\delta,\\
    \psi^{ab}\equiv&\, \Phi^{\alpha\beta\gamma\delta}\, {e^a}_\alpha \,n_\beta \,{e^b}_\gamma \,n_\delta,
\end{align}
and rewrite the symplectic potential as 
\begin{equation}
\begin{split}
    \Theta_g=\int_{\partial\man}\left[\left(K_{cd}\phi^{acbd}-2\mathcal D_c\varphi^{acb}-(\nabla_n+K)\psi^{ab}\right)\delta h_{ab}+2\psi^{ab}\delta K_{ab}\right],
    \label{symplectic potential general}
\end{split}
\end{equation}
where $\mathcal D_a$ is the covariant derivative compatible with the boundary metric and $\nabla_n\psi^{ac}$ is a shorthand for $e^a_\mu \, n_\nu \, e^c_\rho \, n_\sigma \, \nabla_n\Phi^{\mu\nu\rho\sigma}$. Whereas we will use Dirichlet boundary conditions for the metric tensor throughout this work, we chose to keep the associated momentum (i.e.~the term in round brackets) explicit for future investigations.
The above form of the symplectic potential suggests which are the natural boundary conditions to be imposed.

\subsection{The problem: lack of resilience of the bulk-boundary matching}
\label{sec:DDBC}

The simplest possibility one can envision, as put forward in~\cite{Sendouda:2011hq}, is to impose Dirichlet boundary conditions on both the metric and the double normal projection of $\Phi$
\begin{equation}
    \label{Dirichlet-Dirichlet b.c.}
    \delta h_{ab}=\delta\psi^{ab}=0 \qquad \text{on}\;\;\partial\man.
\end{equation}
By doing so, one can easily guess that the needed boundary action takes the form of a generalized Gibbons-Hawking-York boundary term
\begin{equation}
        S^{'\prtl}[g,\Phi]=-2\int_{\partial\man}\psi^{ab} K_{ab},
\label{GHY for Dirichlet}
\end{equation}
up to terms that depend only on $h_{ab}$, $\psi^{ab}$, and their derivatives along the boundary.
In this case, the bulk-boundary matching refers to this action coming with a factor $-2$ or, better, to the particular combination of bulk coefficients that enter this term through the on-shell value of $\psi^{ab}$, that is\footnote{We remind the reader that the subscript $\mathbb T$ instructs us to take the traceless part of a symmetric tensor, cfr. footnote~\ref{Melmed poly}.}
\begin{equation}
    \label{lagrange os}
    \psi^{ab}=-2\alpha_0 (\rho-2\pois) h^{ab}+2\beta_0({}_\mathbb{T}\rho^{ab}+{}_\mathbb{T}\pois^{ab})+2\gamma_0(2\rho^{ab}-\rho h^{ab}).
\end{equation}
In the formula above we introduced for convenience the following tangential tensors obtained by (normal or tangential) projections of the Riemann tensor indices
\begin{align}
    \rho_{ab}&\equiv -R_{\mu\nu\rho\sigma} h^{\nu\sigma}e_a^\mu e^b_\rho,\qquad \rho\equiv h^{ab}\rho_{ab},\\
    \Sigma_{ab}&\equiv R_{\mu\nu\rho\sigma}n^\nu n^\sigma e_a^\mu e_b^\rho,\qquad \Sigma\equiv h^{ab}\Sigma_{ab}.
\end{align}
Notice that the Gauss-Codazzi relation let us write the tensor $\rho_{ab}$ in terms of the boundary intrinsic and extrinsic geometry, whereas, to write the tensor $\pois_{ab}$, one needs more than just the local embedding of the boundary. Replacing the expression~\eqref{lagrange os} in \eqref{GHY for Dirichlet} and using the identities
\begin{equation}
    (\rho-2\Sigma)h^{ab} K_{ab}=R K,\qquad ({}_\mathbb T \rho^{ab}+{}_\mathbb T \pois^{ab})\,{}_\mathbb T K_{ab}=2U,
\end{equation}
we get the explicit expression of the boundary action in terms of boundary operators
\begin{equation}
\label{GHY for Dirichlet expanded}
    S^\prtl_0[g]=\int_{\partial\man}\pqty{4\alpha_0 R K-8\beta_0 U
    -\gamma_0\pqty{B_3+16\det K}}.
\end{equation}
Comparing these terms with the general parametrization of the quadratic action boundary term \eqref{QG action boundary}, we can read the matching that the boundary conditions~\eqref{Dirichlet-Dirichlet b.c.} entail\footnote{We remind the reader of the following identity between cubic polynomials
\begin{equation}
    \det K=\frac{1}{3}G_3-\frac{1}{6}G_2 K+\frac{1}{27}K^3.
\end{equation}}
\begin{equation}
\label{matching before improvement}
S^{\prtl}_0\rightarrow
\begin{cases}
    {}^1a_0=-\dfrac{16}{3}\gamma_0,\\
    {}^2a_0=\dfrac{8}{3}\gamma_0,\\
    {}^3a_0=-\dfrac{16}{27}\gamma_0,\\
    {}^1b_0=-\gamma_0,\\
    {}^2b_0=0,\\
    {}^1c_0=-8\beta_0,\\
    {}^2c_0=4\alpha_0,\\
    {}^3c_0=0.
\end{cases}
\end{equation}

We now investigate whether this matching is preserved by the renormalization procedure. As we extensively discussed in Sec.~\ref{sec:HK}, once we integrate out matter, the coefficients of dimension $4$ operators obtain logarithmic corrections, whose precise form is dictated by the heat kernel coefficient $\mathfrak a_4$~\eqref{aD4 coefficients} or~\eqref{aR4 coefficients}. We thus have
\begin{equation}
    \label{bulk reno}
    \begin{split}
    &\text{In the bulk}\\
    \alpha_\text{ren}&=\alpha_0-\frac{(\xi-\xi_c)^2}{2}\frac{1}{16\pi^2}\ln{\frac{\epsilon}{\mu}},\\
    \beta_\text{ren}&=\beta_0-\frac{1}{120}\frac{1}{16\pi^2}\ln{\frac{\epsilon}{\mu}},\\
    \gamma_\text{ren}&=\gamma_0+\frac{1}{360}\frac{1}{16\pi^2}\ln{\frac{\epsilon}{\mu}},
    \end{split}
\end{equation}
\begin{equation}
    \label{boundary reno}
    \begin{split}
    &\text{In the boundary (Dirichlet b.c.)}\\
        {}^1a_\text{ren}&={}^1a_0-\frac{2}{35}\frac{1}{16\pi^2}\ln{\frac{\epsilon}{\mu}}\,,\\
        {}^2a_\text{ren}&={}^2a_0,\\
        {}^3a_\text{ren}&={}^3a_0,\\
        {}^1b_\text{ren}&={}^1b_0-\frac{1}{360}\frac{1}{16\pi^2}\ln{\frac{\epsilon}{\mu}}\,,\\
        {}^2b_\text{ren}&={}^2b_0,\\
        {}^1c_\text{ren}&={}^1c_0+\frac{1}{15}\frac{1}{16\pi^2}\ln{\frac{\epsilon}{\mu}}\,,\\
        {}^2c_\text{ren}&={}^2c_0-\frac{\xi-\xi_c}{3}\frac{1}{16\pi^2}\ln{\frac{\epsilon}{\mu}}\,,\\
        {}^3c_\text{ren}&={}^3c_0+\frac{\xi-\xi_c}{2}\frac{1}{16\pi^2}\ln{\frac{\epsilon}{\mu}}\,,
    \end{split}\qquad
    \begin{split}
    &\text{In the boundary (Robin b.c.)}\\
        {}^1a_\text{ren}&={}^1a_0-\frac{2}{45}\frac{1}{16\pi^2}\ln{\frac{\epsilon}{\mu}}\,,\\
        {}^2a_\text{ren}&={}^2a_0+\frac{2(\eta-\eta_c)}{15}\frac{1}{16\pi^2}\ln{\frac{\epsilon}{\mu}}\,,\\
        {}^3a_\text{ren}&={}^3a_0+\frac{4(\eta-\eta_c)^3}{3}\frac{1}{16\pi^2}\ln{\frac{\epsilon}{\mu}}\,,\\
        {}^1b_\text{ren}&={}^1b_0-\frac{1}{360}\frac{1}{16\pi^2}\ln{\frac{\epsilon}{\mu}}\,,\\
        {}^2b_\text{ren}&={}^2b_0-\frac{2(\eta-\eta_c)}{45}\frac{1}{16\pi^2}\ln{\frac{\epsilon}{\mu}}\,,\\
        {}^1c_\text{ren}&={}^1c_0+\frac{1}{15}\frac{1}{16\pi^2}\ln{\frac{\epsilon}{\mu}}\,,\\
        {}^2c_\text{ren}&={}^2c_0-(\xi-\xi_c)(2\rb-\rb_c)\frac{1}{16\pi^2}\ln{\frac{\epsilon}{\mu}}\,,\\
        {}^3c_\text{ren}&={}^3c_0-\frac{\xi-\xi_c}{2}\frac{1}{16\pi^2}\ln{\frac{\epsilon}{\mu}}\,,
    \end{split}
\end{equation}
The reader will notice immediately that some coefficients do not renormalize if the scalar field is conformally coupled to the curvature ($\xi=\xi_c$) and its boundary condition are Weyl invariant (either Dirichlet or Robin with $\eta=\eta_c$). This is because the $\mathfrak{a}_4$ coefficient in the heat kernel of a conformally invariant differential operator is conformally invariant itself~\cite{Dowker:1978md} so, in that case, only the coefficients corresponding to conformally invariant operators ---~namely $\beta$, $\gamma$, ${}^1a_0$, ${}^1b_0$, ${}^1c_0$~--- are renormalized at the $\ln\epsilon$ level.

If we thus select the conformal values of the couplings, we see that one can preserve the bulk-boundary matching \eqref{matching before improvement} for all $b$- and $c$-type boundary constants (the last five of \eqref{matching before improvement}). This is not the case for the $a$-type couplings (the first three). 
Even worse, there is no way to preserve the bulk-boundary matching for all of them: neither the minimal case nor the idea of setting $\eta=\xi$, which seemed to yield a resilient matching of Newton's constants (modulo the previously discussed potential obstruction), appear to work here.

\section{Moving forward: an improved boundary action}
\label{sec:improved}

In the previous section we found that starting with the generalized GHY term~\eqref{GHY for Dirichlet}, there is apparently no way to preserve the bulk-boundary matching upon renormalization.
However, as already mentioned in~\cite{Deruelle:2009zk}, it might not be always correct to use the generalized Gibbons-Hawking term as in~\eqref{GHY for Dirichlet}. We shall now expand on that comment by showing that imposing Dirichlet boundary conditions on $\psi^{ab}$ is not as natural as one might think. To illustrate this point, let us start by reviewing a simple extension of Einstein's gravity which does not modify the dynamical content of the theory.

In the construction of Lanczos-Lovelock theories~\cite{Lanczos:1938sf, Lovelock:1971yv, Padmanabhan:2013xyr}, one considers the most general action built out of the Riemann tensor which yields differential equations for the metric in $d$-dimensions which are second-order despite the appearance of higher-curvature operators. The result is a sum of terms, polynomial in the curvature tensor, multiplied by arbitrary coefficients $\varsigma_p$
\begin{equation}
    S^{LL}_{(d)}=\int_\man  \sum_p\varsigma_p L_p,\qquad L_p=\frac{1}{2^p}\delta^{\mu_1\dots \mu_{2p}}_{\nu_1\dots \nu_{2p}}R^{\nu_1 \nu_2}_{\mu_1 \mu_2}\cdots R^{\nu_{2p-1}\nu_{2p}}_{\mu_{2p-1}\mu_{2p}}.
\end{equation}
The symbol $\delta^{\mu_1\dots \mu_{2p}}_{\nu_1\dots \nu_{2p}}$ is the Kronecker symbol of order $2p$, given by the determinant of a matrix whose $i,j$ entries are $\delta^{\mu_i}_{\nu_j}$. In any even dimension $d$, the sum stops at $p=\lfloor{d/2}\rfloor$. Moreover, such a term is topological for a manifold of dimension $d=2p$, so it starts to contribute to the equations of motion only for manifolds with $d>2p$.

For our discussion, we take $d=4$, so that the sum contains just three operators: $L_0=1$, $L_1=-R$ and $L_2=R^2-4R_{\mu\nu}R^{\mu\nu}+R_{\mu\nu\rho\sigma}R^{\mu\nu\rho\sigma}=E_4$. The Lanczos-Lovelock theory in $d=4$ is just Einstein-Gauss-Bonnet gravity, in which one supplements the Einstein-Hilbert term with the Gauss-Bonnet invariant. Being topological, the latter does not contribute to the equations of motion, which remain those of GR.

\subsection{A lesson from Gauss-Bonnet gravity}
\label{sec:LL}

Let us then see what would happen if we blindly apply the prescription \eqref{GHY for Dirichlet} to $4$d Lanczos-Lovelock theory. Given that the bulk action is $S_{\text{cl}}^{\text{bulk}}=\int_\man(\varsigma_0-\varsigma_1 R+\varsigma_2 E_4)$, we should write the boundary term as
\begin{equation}
    S_{\text{cl}}^\prtl=-\int_{\partial\man}\pqty{\varsigma_1 K+\varsigma_2(B_3+16\det K)}\,,
\end{equation}
where $B_3$ was defined in Eq~\eqref{eq:Chern}. However, this conclusion is in conflict with the literature on Gauss-Bonnet gravity~\cite{Myers:1987yn, Davis:2002gn, Gravanis:2002wy, Deruelle:2003ps}. As a matter of fact, the Euler theorem tells us that the boundary term that compensates for the bulk variation of $E_4$ is just (minus) $B_3$ since
\begin{equation}
    \int_\man E_4-\int_{\partial\man}B_3=32\pi^2\chi[\man],
\end{equation}
with $\chi[\man]$ the Euler characteristic of $\man$. As long as the metric is held fixed on the boundary, this combination is manifestly invariant, so its contribution to the variational principle, while trivial, is well-defined.

The problem with our prescription is that, by construction, the action $S_{\text{cl}}^{LL}$ yields equations of motion for the metric which are second-order (the same of GR), hence they only propagate the standard graviton. The introduction of auxiliary fields in this case is superfluous and no additional boundary condition should be required to make the boundary-value problem well-defined.
What happens is that, for this theory, all the components of $\psi^{ab}$ are determined by the Gauss-Codazzi relations in terms of the intrinsic and extrinsic curvature of $\partial\man$\footnote{In the case of GR, the tensor $\Phi^{\mu\nu\rho\sigma}$ is proportional to a combination of the metric tensor, which makes it clear that this field has no independent dynamical content. Accordingly, the presence of $R$ in the bulk action does not change the conclusion of this section.}.
Since the intrinsic geometry of the boundary is already fixed, requiring Dirichlet boundary conditions on $\psi^{ab}$ would be equivalent to fixing the extrinsic curvature too. However, this would be a statement ---~a Neumann boundary condition~---  about the metric tensor, i.e.~a different degree of freedom. The conclusion is that in $4$d Einstein-Gauss-Bonnet theory, the auxiliary field $\Phi^{\mu\nu\rho\sigma}$ has no dynamical content (as in GR), so it is not natural, nor necessary, to impose boundary conditions on it.

\subsection{An improved boundary term for quadratic gravity}
\label{sec:IGH}

Learning from the above particular case, we now implement a different prescription that works for a generic action with quadratic curvature operators.
First, we add and subtract the topological term $E_4$ from the Lagrangian to get $f(\text{Riem})=(f(\text{Riem})-\gamma_0 E_4)+\gamma_0 E_4\equiv\tilde f(\text{Riem})+\gamma_0 E_4$. This splitting induces a similar one in $\Phi^{\mu\nu\rho\sigma}$ and in $\psi^{ab}$.
\begin{equation}
    f=\tilde f+\gamma_0 E_4\quad\rightarrow\quad\Phi^{\mu\nu\rho\sigma}=\tilde \Phi^{\mu\nu\rho\sigma}+\bar\Phi^{\mu\nu\rho\sigma}\quad\rightarrow \quad\psi^{ab}=\tilde\psi^{ab}+\bar{\psi}^{ab}.
\end{equation}
One could refer to the first (tilded) terms as dynamical, and to the second (barred) terms as topological. To be specific, we have
\begin{equation}
\label{dyn vs top splitting}
    \tilde\psi^{ab}=-2\alpha_0 (\rho-2\pois) h^{ab}+2\beta_0({}_\mathbb{T}\pois^{ab}+{}_\mathbb{T}\rho^{ab}),\qquad \bar\psi^{ab}=2\gamma_0(2\rho^{ab}-\rho h^{ab}).
\end{equation}
If we perform this split in the symplectic potential
\eqref{symplectic potential general}
\begin{equation}
    \Theta_g=\int_\man\pqty{(\dots)\delta h_{ab}+2\tilde\psi^{ab}\delta K_{ab}+2\bar\psi^{ab}\delta K_{ab}}
    \label{eq:splitsympl}
\end{equation}
it is clear that we should not use the Leibniz rule in the last term to extract a total variation because, as we learned in the last section, imposing conditions on $\bar\psi^{ab}$ is equivalent to imposing conditions on $K_{ab}$. Actually, the last term in Eq.~\eqref{eq:splitsympl} is already a total variation, as one can check by explicit calculation
\begin{equation} 2\bar\psi^{ab}\delta K_{ab}=\gamma_0 \delta B_3+\mathcal O(\delta h_{ab}).
\end{equation}
Therefore we see that the generalized GHY Lagrangian should contain $B_3$ alone.

If we now assume the following set of Dirichlet--Dirichlet boundary conditions
\begin{equation}
    \delta h_{ab}=\delta\tilde\psi^{ab}=0 \qquad \text{on}\;\;\partial\man,
\end{equation}
we see that the correct boundary action is
\begin{equation}
    \label{GHY for Dirichlet improved}
     S^{'\prtl,\,\text{impr}}_0[g,\Phi]=-\int_{\partial\man}\pqty{2\tilde\psi^{ab}K_{ab}+\gamma_0 B_3}.
\end{equation}
Apart from reducing to the known result for Einstein-Gauss-Bonnet gravity, this action has the advantage of implementing boundary conditions only on the true dynamical d.o.f. of the theory. Indeed, if we split $\tilde\psi^{ab}$ in trace and traceless part
\begin{equation}
    S^{'\prtl,\,\text{impr}}_0[g,\Phi]=-\int_{\partial\man}\pqty{\frac{2}{3}\tilde\psi K+2{}_\mathbb{T}\tilde\psi^{ab}{}_\mathbb{T}K_{ab}+\gamma_0 B_3},
\end{equation}
we see that the generalized GHY action has a particularly simple structure, where the first two terms nicely correspond to the two propagating modes (the scalar and the traceless tensor) of the $\tilde f(\text{Riem})$ theory. The third term combines with the $\gamma_0 E_4$ term in the bulk to produce the Euler characteristic and leave no other term that survives in the flat spacetime limit. Alternatively setting $\alpha_0$ and $\beta_0$ to zero cancels the first and the second term, as one can see from the first of~\eqref{dyn vs top splitting}, suggesting that there are no other possible reductions.

The improved GHY boundary action implements a new bulk-boundary matching
\begin{equation}
\label{matching after improvement}
S^{\prtl,\,\text{impr}}_0\rightarrow
\begin{cases}
        {}^1a_0=0,\\
        {}^2a_0=0,\\
        {}^3a_0=0,\\
        {}^1b_0=-\gamma_0,\\
        {}^2b_0=0,\\
        {}^1 c_0=-8\beta_0,\\
        {}^2c_0=4\alpha_0,\\
        {}^3c_0=0,
    \end{cases}
\end{equation}
which differs from \eqref{matching before improvement} in the first three conditions.

Again, looking at~\eqref{bulk reno} and~\eqref{boundary reno}, we see that the process of integrating out a scalar field with conformal couplings leads to a renormalization of the $\beta_0,\gamma_0,{}^1a_0,{}^1b_0,{}^1c_0$ coefficients only, leaving $\alpha_0, {}^2a_0, {}^3 a_0, {}^2 b_0,{}^2c_0, {}^3c_0$ unchanged. Therefore, the matching conditions involving the latter are trivially preserved. As far as the former are concerned, the reader can check that the matching of ${}^1 b_0$ with $\gamma_0$ and ${}^1 c_0$ with $\beta_0$ are preserved too. Unfortunately, the fly in the ointment is the coefficient ${}^1a$ (multiplying $G_3$ in the boundary action \eqref{QG action boundary}), which necessarily acquires a non-zero contribution, breaking the corresponding bulk-boundary matching which wants it to be $0$. As a side comment, let us stress that the minimal and $\rb=\xi$ cases are still unfeasible as they spoil more matching conditions than the conformal one.

\subsection{The $G_3$ term}

We can summarize the outcome of the previous section by saying that the identification of a proper boundary action, isolating the truly dynamical degrees of freedom of the theory from the topological contributions, leads to an almost complete preservation of the bulk-boundary matching of quadratic couplings in the presence of a conformally coupled scalar field. In particular, it fixes the problem with the ${}^2a$ and ${}^3a$ coefficients, but not with ${}^1 a$. Nevertheless, it fails to do so in an interesting way, on which we shall now comment

It is thus clear that the reason for our shortcoming is that the GHY term does not contain the $G_3$ operator \eqref{eq:Gi}. At the current stage, we have not yet understood the meaning of this object in the context of the variational principle, or the physical reasons why one should expect it to appear in the boundary action. Nevertheless, we can offer a reason why this term might not be so problematic.

Let us start by pointing out that $G_3$ trivially vanishes if the extrinsic curvature tensor has no trace-free component, i.e.~$K_{ab}=1/3 K h_{ab}$. Under this assumption, the only possible contribution of the boundary extrinsic curvature to the matter effective action reduces to $K^3$, which, being conformally variant, cannot be present if we restrict to conformally invariant matter fields, as we do. Since $G_3$ does not enter the boundary action, we do not have to worry about the matching of its corresponding coefficient.

The condition $K_{ab}=1/3 K h_{ab}$ has been frequently used in the renormalization of pure gravity, either perturbative or within the exact functional renormalization group~\cite{Jacobson:2013yqa,Falls:2017cze}. As a matter of fact, it plays a role in the formulation of the boundary value problem for gravitons. It has been first identified in~\cite{Moss:1996ip} as a way to get simple BRST-invariant local boundary conditions (see also~\cite{Luckock:1990xr,Esposito:1997wt}), and it also allows to preserve the strong ellipticity of the problem without introducing tangential derivatives in the boundary conditions~\cite{Avramidi:1997sh,Moss:2013vh}.
The fact that highly-energetic graviton loops will eventually be integrated out appears to us as a good motivation to constrain the form of $K_{ab}$ in such a way.

Finally, let us notice that this problem is not due to the specific kind of matter we are considering. Indeed, the $\mathfrak{a}_4$ coefficient of the heat kernel for a generic set of conformally invariant matter fields (conformally coupled massless scalars, Weyl fermions, Maxwell fields) can be written as~\cite{Fursaev:2015wpa}
\begin{equation}
    \mathfrak{a}_4=\frac{4\pi^2\, p}{45}\chi[\man]+\frac{q}{120}\int_{\man}C^2-\int_{\partial\man}\pqty{\frac{q_1}{15}U-\frac{2t}{35}G_3},
\end{equation}
with specific values of the $p$, $q$, $q_1$, $t$ charges. Since the operators $E_4$ and $B_3$ enter this coefficient only through $\chi[\man]$, the matching between ${}^1 b$ and $\gamma$ will not be spoiled. Likewise, since $q=q_1$ for all relevant matter fields, the matching between ${}^1 c$ and $\beta$ will not be spoiled either. On the other hand, the fact that the $t$ charge is positive for all fields --- it is $1$ for Dirichlet scalar fields, $7/9$ for conformal Robin scalar fields, $5$ for Weyl fermions and $8$ for Maxwell fields --- means that there is no particular combination of the matter content that would make the problem disappear by yielding a net vanishing loop contribution to ${}^1 a$. This seems to imply that the $G_3$ term is universal and, as such, its eventual removal should be based on very general grounds.

\section{Discussion}
\label{sec:CN}

In this paper, we have explored the issue of whether renormalization preserves or not the bulk-boundary matching that we expect at the bare classical level to have a well-posed variational principle. In particular, we focused on the coefficients of quadratic curvature operators, which had not been previously discussed.
To do so, we first discussed the appearance of additional degrees of freedom in higher-curvature theories, whose presence affects the structure of the boundary action. \textit{En passant}, we addressed the subtlety noticed in~\cite{Deruelle:2003ps} and provided a different prescription to construct the GHY boundary term. Shortly explained, it might happen that the theory has a particular combination of couplings that make part of the expected d.o.f. non-dynamical. In this case, the determination of the form of the required boundary term is not straightforward and should be dealt with on a case-by-case basis.
This is what happens in Einstein-Gauss-Bonnet (a Lanckzos-Lovelock theory), in which none of the expected extra degrees of freedom actually show up. 
Once we had determined the form of the boundary action implementing the improved boundary conditions, we were able to infer the bulk-boundary matching. Finally, we explored how renormalization affects each side of the matching, either preserving or spoiling it.

Our study reveals that conformal invariance plays a crucial role in the resilience of the matching. While we presented the problem in its general form, we were later forced to consider only the set of couplings that make the dynamics and the boundary conditions invariant under Weyl rescaling. Let us briefly comment on this.

The renormalization of dimensionless couplings, as those multiplying the quadratic curvature operators, cannot be affected by the mass of the field for dimensional reasons. Hence we can focus on massless fields without loss of generality. We know that massless vector bosons in $4$d and massless fermions are Weyl invariant. Additionally, it has been shown in~\cite{Sonego:1993fw} that any other choice than $\xi=\xi_c$ leads to a conflict with the equivalence principle in the form of a violation of Huygens's principle from the timelike tails of the massless scalar field $2$-point function. Finally, there is circumstantial evidence in the literature that the RG flow of a scalar field coupled to gravity naturally leads to the conformal value in the IR \cite{DeSimone:2008ei, PhysRevD.25.1019}. Hence, what may appear to be a limitation of our analysis can be seen as a natural expectation of what happens at the low energies we are implicitly considering, consistently with the regime of validity of the effective action for the gravitational field that we constructed.

The careful reader would have surely noticed that the idea of setting $\xi=\xi_c$ and $\eta=\eta_c$, which makes most of the matching resilient upon renormalization, contrasts with the condition $\eta=\xi$ that was found in \cite{Jacobson:2013yqa} to ensure a consistent matching of Newton's constants. Nonetheless, we believe that the here proposed setting is supported and to be preferred by at least three arguments. First, the renormalization of Newton's constant is plagued by regularization and gauge dependencies. In order to construct observables, Newton's constant has to be combined with some other non-universal and dimensionful parameters into a quantity with scheme-independent running. One might envision that the renormalization of these other parameters is $\eta$ dependent too and in such a way that, when combined with the boundary Newton's constant, the two mismatches cancel out. 
Second, had we started from the very beginning with a regularization procedure that does not introduce a mass scale, like the zeta-function regularization, the effective action would have contained only the logarithmic divergences so that, for Weyl invariant matter fields, Newton's constant would have been unaffected by the renormalization procedure.
Third, as we discussed in Sec.~\ref{sec:D2} and shown in Appendix~\ref{app:scalar}, the Robin boundary condition needed to get $\eta=\xi$ cannot be implemented as straightforwardly as those for $\eta=2\xi$, which is compatible with the conformal values.

In conclusion, we hope that this investigation will help to shed some new light on an undeservedly neglected issue in quantum field theory in curved spacetime. The main lessons we can draw from it is that, as usual when dealing with generalized theories of gravity, the identification of the truly dynamical field in the theory is paramount and that what spoils the full bulk-boundary matching is a universal term, whose removal we suspect to rest on more fundamental arguments than those provided here. 
We hope that the present work will stimulate further investigations on the issue and lead to its definitive resolution.

\appendix
\section*{Appendix}
\section{Geometry of the bulk and the boundary}
\label{app:GI}

This appendix sets the notation and contains much  and it is mainly drawn from \cite{Vassilevich:2003xt}, whose convention we mostly follow\footnote{The only difference is that we use $n^\mu$ to denote the outward pointing normal to $\partial \man$, whence the plus sign in the definition of the extrinsic curvature \eqref{extrinsic curvature}}.
We assume this manifold to be equipped with a positive definite metric $g$ and to have a non-empty boundary $\partial\man$. The metric allows us to define a unique connection whose covariant derivative we denote by $\nabla_\mu$.
Out of these objects, we can construct the curvature tensor
\begin{equation}
    R^\mu{}_{\nu\rho\sigma}=\partial_\sigma \Gamma^\mu_{\nu\rho}-\partial_\rho \Gamma^\mu_{\nu\sigma}+\Gamma^\lambda_{\nu\rho}\Gamma^\mu_{\lambda\sigma}-\Gamma^\lambda_{\nu\sigma}\Gamma^\mu_{\lambda\rho}
\end{equation}
and its contractions $R_{\mu\nu}\equiv R^\lambda{}_{\mu\nu\lambda}$, $R\equiv R^\lambda_\lambda$. Note that, with this convention, the curvature is defined by the commutator as $[\nabla_\rho, \nabla_\sigma]V^\mu=-R^\mu{}_{\nu\rho\sigma}V^\nu$.

The presence of the boundary $\partial\man$ brings about more structure and thus the possibility of building more invariants. Once we have the map that embeds the boundary in the spacetime, we can pull-back the metric $g$ to $\partial\man$ to get the induced metric
\begin{equation}
    h_{ab}=g_{\mu\nu}e^\mu_a e^\nu_b,
\end{equation}
and construct a unit vector $n$ which is normal to $\partial\man$ and points outward. This leads to a natural decomposition of the tangent space of $\man$ into the tangent space of $\partial\man$ and its orthogonal complement which is expressed by the split
\begin{equation}
    g_{\mu\nu}=h_{\mu\nu}+n_\mu n_\nu,
    \label{completeness}
\end{equation}
where we denoted by $h_{\mu\nu}$ the induced metric push-forward into the boundary (image). The covariant derivative $\nabla_\mu$ splits accordingly. 
However, the derivative of a vector $V^\alpha$ tangent to the boundary (i.e. $n_\alpha V^\alpha=0$) along the boundary itself does not remain tangent but it acquires a normal component according to
\begin{equation}
    e^\nu_b \nabla_\nu V^\alpha=\mathcal D_b V^a e^\alpha_a-V^a K_{ab} n^\alpha.
\end{equation}
In the last expression, we introduced $V^a$ as the pull-back of $V^\alpha$ to the boundary, $\mathcal D_a$ as the covariant derivative compatible with $h_{ab}$ and the extrinsic curvature tensor
\begin{equation}
    K_{ab}\equiv e^\alpha_a e^\beta_b \nabla_\beta n_\alpha,
    \label{extrinsic curvature}
\end{equation}
which then measures the difference between the tangential projection of the bulk covariant derivative and the tangential boundary covariant derivative.

Splitting all the indices of the (bulk) Riemann tensor, one can write
\begin{equation}
\begin{split}
    R_{\alpha\beta\gamma\delta}=\rho_{abcd}e^a_\alpha e^b_\beta e^c_\gamma e^d_\delta+\pi_{abc}e^a_\alpha e^b_\beta e^c_\gamma n_\delta+\pois_{ac}e^a_\alpha e^c_\gamma n_\beta n_\delta+\text{symm.}
    \label{horizontal split of R}
\end{split}
\end{equation}
where we introduced the contractions
\begin{align}
    \rho_{abcd}&\equiv R_{\alpha\beta\gamma\delta}e^\alpha_a e^\beta_b e^\gamma_c e^\delta_d,\\
    \pi_{abc}&\equiv R_{\alpha\beta\gamma\delta}e^\alpha_a e^\beta_b e^\gamma_c n^\delta,\\
    \pois_{ac}&\equiv R_{\alpha\beta\gamma\delta}e^\alpha_a n^\beta e^\gamma_c n^\delta,
\end{align}
which are tensors in the tangent space of $\partial\man$, and we wrote symm. in place of those terms that can be obtained from the others using the symmetries of $R_{\alpha\beta\gamma\delta}$.
Likewise, we have
\begin{equation}
    R_{\alpha\beta}=\rho_{ab}e_\alpha^a e^b_\beta-2\pi_{a} e^a_{(\alpha} n_{\beta)} -\pois_{ab} e^a_\alpha e^b_\beta -\pois n_\alpha n_\beta,
\end{equation}
and
\begin{equation}
    R=\rho-2\pois,
\end{equation}
with $\rho_{ab}\equiv h^{cd}\rho_{cabd}$, $\rho=h^{ab}\rho_{ab}$, $\pi_a\equiv h^{bc}\pi_{bac}$ and $\pois\equiv\pois_{ab}h^{ab}$. The geometrical meaning of these projections of the curvature tensor is related to the boundary intrinsic and extrinsic curvature by the famous Gauss-Codazzi equations.
In formulas,
\begin{gather}
    \rho_{abcd}=\mathcal R_{abcd}+K_{ac}K_{bd}-K_{ad}K_{bc},\\
    \pi_{abc}=\mathcal D_b K_{ac}-\mathcal D_a K_{bc}.
\end{gather}
On the other hand, it is known that the $\Sigma_{ab}$ tensor cannot be written solely in terms of the surface intrinsic and extrinsic geometry.

\section{Scalar field action}
\label{app:scalar}

Throughout the paper, we only used the scalar field to produce an effective contribution to the gravitational action. Here, we shall give a more detailed description of its dynamics.

We take the bulk part of the scalar field action to be given by
\begin{equation}
    S^{\text{bulk}}_m=\frac{1}{2}\int_{\man}\pqty{\nabla_\mu\phi\,\nabla^\mu\phi+\xi R\phi^2}.
\end{equation}
Varying the action $S^{\text{bulk}}_m$ with respect to the dynamical fields $\phi$ and $g_{\mu\nu}$ yields the equation of motion for the scalar field and the source for the equations of motion for the metric ---~which come from the variation of $S_\text{cl}[g]$ in Eq.~\eqref{partition function}~---, up to a total derivative that defines the symplectic potential
\begin{equation}
    \begin{split}
    \delta S^{\text{bulk}}_m&=\int_{\man}\biggl[\frac{1}{2}\delta g^{\mu\nu}\pqty{\nabla_\mu\phi \nabla_\nu\phi+\xi R_{\mu\nu}\phi^2-\frac{1}{2}g_{\mu\nu}(\nabla_\sigma\phi\nabla^\sigma\phi+\xi R\phi^2)}+\\
    &\qquad\quad+\nabla_\mu\phi\nabla^\mu\delta \phi+\xi R\phi\,\delta\phi+\frac{1}{2}\xi g^{\mu\nu}\delta R_{\mu\nu}\phi^2\biggr]=\\
    &=\int_{\man}\biggl[\frac{1}{2}\delta g^{\mu\nu}\pqty{\nabla_\mu\phi \nabla_\nu\phi+\xi R_{\mu\nu}\phi^2-\frac{1}{2}g_{\mu\nu}(\nabla_\sigma\phi\nabla^\sigma\phi+\xi R\phi^2)}+\\
    &\qquad\quad+\delta\phi(-\Box+\xi R)\phi+\xi g^{\mu\nu} \delta g_{\mu[\sigma}g^{\sigma\rho}\nabla_{\nu]}\nabla_\rho\phi^2\biggr]+\\
    &\quad+\int_{\partial\man}\pqty{n_\mu \delta\phi\nabla^\mu\phi+\xi g^{\mu\nu}n^\sigma \nabla_{[\nu}\delta g_{\sigma]\mu}\phi^2-\xi g^{\mu\nu}\delta g_{\mu[\sigma}n_{\nu]}\nabla^\sigma\phi^2}=\\
    &=\int_{\man}\pqty{\delta\phi(-\Box+\xi R)\phi+\frac{1}{2}\delta g^{\mu\nu}T_{\mu\nu}}+\int_{\partial\man} \theta_m.
    \end{split}
\end{equation}
In the last line, we identified the stress-energy tensor of matter
\begin{equation}
    T_{\mu\nu}=\nabla_\mu\phi \nabla_\nu\phi+\xi R_{\mu\nu}\phi^2-\xi(\nabla_\mu \nabla_\nu-g_{\mu\nu}\nabla^2)\phi^2 -\frac{1}{2}g_{\mu\nu}\pqty{\nabla_\sigma\phi\,\nabla^\sigma\phi+\xi R\phi^2},
\end{equation}
which sources the metric field equations as $\mathcal E_{\mu\nu}=T_{\mu\nu}$, and the symplectic potential
\begin{equation}
    \Theta_m=\int_{\partial\man}\theta_m=\int_{\partial\man}\pqty{\delta\phi \nabla_n\phi+\frac{1}{2}\xi\pqty{h^{ab}\nabla_n \phi^2-K^{ab}\phi^2}\delta h_{ab}-\xi\phi^2\delta K},
\end{equation}
that we wrote in terms of variations of the induced metric and the extrinsic curvature of $\partial\man$, dropping a total horizontal derivative.
If we integrate the last term, we obtain
\begin{equation}
\label{scalar field symplectic potential}
    \Theta_m=\int_{\partial\man}\pqty{(\nabla_n\phi+2\xi K\phi)\delta \phi+\frac{1}{2}\xi(h^{ab}\nabla_n-K^{ab}+K h^{ab})\phi^2\delta h_{ab}}-\delta\int_{\partial\man}(\xi K \phi^2),
\end{equation}
from which we recognize the need to add a $\phi$-dependent Gibbons-Hawking-York boundary term. Moreover, we can read the momenta conjugated to the boundary phase space variables $(\phi, h_{ab})$, namely
\begin{equation}
    p=\sqrt{h}(\nabla_n\phi+2\xi K\phi),\quad \Pi^{ab}=\frac{\sqrt{h}}{2}\xi(h^{ab}\nabla_n-K^{ab}+K h^{ab})\phi^2.
\end{equation}

In order to obtain a well-defined variational principle, we have to make sure that the total boundary variation of the action vanishes under the correct boundary conditions. To do so, we are free to add a boundary term to the action. For example, it is easy to see how one would implement the Dirichlet-Dirichlet boundary conditions
\begin{equation}
    \delta\phi=0,\quad \delta h_{ab}=0 \qquad\text{on}\;\;\partial\man.
\end{equation}
Indeed, the first two terms in~\eqref{scalar field symplectic potential} vanish so that the appropriate boundary term is just the GHY
\begin{equation}
    S_{m\,(D)}^{\prtl}=\int_{\partial\man}\xi K\phi^2.
\end{equation}
If, on the other hand, we want the scalar field momentum $p$ to be fixed at the boundary, while keeping Dirichlet boundary conditions for the metric, i.e.
\begin{equation}
    \delta(\nabla_n\phi+2\xi K\phi)=0,\quad \delta h_{ab}=0 \qquad\text{on}\;\;\partial\man,
\end{equation}
we can just integrate by part the first term and cancel the total variation by adding $-p\phi$ to the boundary action 
\begin{equation}
    S_{m\,(N)}^{\prtl}=\int_{\partial\man}\pqty{\xi K\phi^2-\phi\nabla_n\phi-2\xi K\phi^2}=-\int_{\partial\man}\pqty{\phi\nabla_n\phi+\xi K\phi^2}.
\end{equation}

However, these are not the only possibilities. As shown in~\cite{Gustafson1998DomainD}, we are free to fix on the boundary any constant linear combination of variables and momenta to obtain the so-called Robin boundary conditions. As far as we are concerned, we will not change the boundary conditions for the metric, so we will only consider the following set of Robin-Dirichlet boundary conditions
\begin{equation}
    \label{general robin}
    \delta(\nabla_n\phi+2\xi K\phi+c\phi)=0,\quad \delta h_{ab}=0 \qquad\text{on}\;\;\partial\man.
\end{equation}
In this case, the correct boundary action is
\begin{equation}
\label{scalar field Robin boundary action}
    S_{m\,(R)}^{\prtl}=\int_{\partial\man}\pqty{\xi K\phi^2-\phi\nabla_n\phi-2\xi K\phi^2-\frac{c}{2}\phi^2}=-\int_{\partial\man}\pqty{\phi\nabla_n\phi+\xi K\phi^2+\frac{c}{2}\phi^2}.
\end{equation}

We now can see why the idea of setting $\eta=\xi$ to construct a resilient matching for Newton's constants faces a problem. If we want to use this construction to implement the needed Robin boundary, i.e. $(\nabla_n+\xi K)\phi=0$, we need to take $c=(\rb-2\xi)K=-\xi K$ in~\eqref{general robin} and~\eqref{scalar field Robin boundary action}. Unfortunately, this is not a constant, and so it will produce an additional, unmatched, term in $\delta K$ when varied with respect to the metric tensor, spoiling the metric variational principle.

\section*{Acknowledgements}

We would like to express our gratitude to Ted Jacobson, Alessio Baldazzi, Laura Donnay, Roberto Percacci, Gloria Odak, Goffredo Chirco and Christopher Eling for stimulating discussion and comments throughout the development of this manuscript. G. Neri especially wishes to thank Beniamino Valsesia for his daily help and support.

% ============ BIBLIOGRAPHY
\bibliography{Biblio}

\providecommand{\href}[2]{#2}\begingroup\raggedright\begin{thebibliography}{10}

\bibitem{Susskind:1994sm}
L.~Susskind and J.~Uglum, \emph{{Black hole entropy in canonical quantum
  gravity and superstring theory}},
  \href{https://doi.org/10.1103/PhysRevD.50.2700}{\emph{Phys. Rev. D}
  {\bfseries 50} (1994) 2700}
  [\href{https://arxiv.org/abs/hep-th/9401070}{{\ttfamily hep-th/9401070}}].

\bibitem{Fursaev:1994ea}
D.V.~Fursaev and S.N.~Solodukhin, \emph{{On one loop renormalization of black
  hole entropy}},
  \href{https://doi.org/10.1016/0370-2693(95)01290-7}{\emph{Phys. Lett. B}
  {\bfseries 365} (1996) 51}
  [\href{https://arxiv.org/abs/hep-th/9412020}{{\ttfamily hep-th/9412020}}].

\bibitem{Frolov:1995xe}
V.P.~Frolov, D.V.~Fursaev and A.I.~Zelnikov, \emph{{Black hole entropy:
  Off-shell versus on-shell}},
  \href{https://doi.org/10.1103/PhysRevD.54.2711}{\emph{Phys. Rev. D}
  {\bfseries 54} (1996) 2711}
  [\href{https://arxiv.org/abs/hep-th/9512184}{{\ttfamily hep-th/9512184}}].

\bibitem{Frolov:1995pt}
V.P.~Frolov, \emph{{Black hole entropy and physics at Planckian scales}},
  {\emph{NATO Sci. Ser. C} {\bfseries 476} (1996) 187}
  [\href{https://arxiv.org/abs/hep-th/9510156}{{\ttfamily hep-th/9510156}}].

\bibitem{Larsen:1995ax}
F.~Larsen and F.~Wilczek, \emph{{Renormalization of black hole entropy and of
  the gravitational coupling constant}},
  \href{https://doi.org/10.1016/0550-3213(95)00548-X}{\emph{Nucl. Phys. B}
  {\bfseries 458} (1996) 249}
  [\href{https://arxiv.org/abs/hep-th/9506066}{{\ttfamily hep-th/9506066}}].

\bibitem{Frolov:1996aj}
V.P.~Frolov, D.V.~Fursaev and A.I.~Zelnikov, \emph{{Statistical origin of black
  hole entropy in induced gravity}},
  \href{https://doi.org/10.1016/S0550-3213(96)00678-5}{\emph{Nucl. Phys. B}
  {\bfseries 486} (1997) 339}
  [\href{https://arxiv.org/abs/hep-th/9607104}{{\ttfamily hep-th/9607104}}].

\bibitem{Frolov:1998vs}
V.P.~Frolov and D.V.~Fursaev, \emph{{Thermal fields, entropy, and black
  holes}}, \href{https://doi.org/10.1088/0264-9381/15/8/001}{\emph{Class.
  Quant. Grav.} {\bfseries 15} (1998) 2041}
  [\href{https://arxiv.org/abs/hep-th/9802010}{{\ttfamily hep-th/9802010}}].

\bibitem{Cooperman:2013iqr}
J.H.~Cooperman and M.A.~Luty, \emph{{Renormalization of Entanglement Entropy
  and the Gravitational Effective Action}},
  \href{https://doi.org/10.1007/JHEP12(2014)045}{\emph{JHEP} {\bfseries 12}
  (2014) 045} [\href{https://arxiv.org/abs/1302.1878}{{\ttfamily 1302.1878}}].

\bibitem{Gibbons:1976ue}
G.W.~Gibbons and S.W.~Hawking, \emph{{Action Integrals and Partition Functions
  in Quantum Gravity}},
  \href{https://doi.org/10.1103/PhysRevD.15.2752}{\emph{Phys. Rev. D}
  {\bfseries 15} (1977) 2752}.

\bibitem{Barvinsky:1995dp}
A.D.~Barvinsky and S.N.~Solodukhin, \emph{{Nonminimal coupling, boundary terms
  and renormalization of the Einstein-Hilbert action and black hole entropy}},
  \href{https://doi.org/10.1016/0550-3213(96)00438-5}{\emph{Nucl. Phys. B}
  {\bfseries 479} (1996) 305}
  [\href{https://arxiv.org/abs/gr-qc/9512047}{{\ttfamily gr-qc/9512047}}].

\bibitem{Becker:2012js}
D.~Becker and M.~Reuter, \emph{{Running boundary actions, Asymptotic Safety,
  and black hole thermodynamics}},
  \href{https://doi.org/10.1007/JHEP07(2012)172}{\emph{JHEP} {\bfseries 07}
  (2012) 172} [\href{https://arxiv.org/abs/1205.3583}{{\ttfamily 1205.3583}}].

\bibitem{Falls:2017cze}
K.~Falls, \emph{{Physical renormalization schemes and asymptotic safety in
  quantum gravity}},
  \href{https://doi.org/10.1103/PhysRevD.96.126016}{\emph{Phys. Rev. D}
  {\bfseries 96} (2017) 126016}
  [\href{https://arxiv.org/abs/1702.03577}{{\ttfamily 1702.03577}}].

\bibitem{Jacobson:2013yqa}
T.~Jacobson and A.~Satz, \emph{{On the renormalization of the Gibbons-Hawking
  boundary term}},
  \href{https://doi.org/10.1103/PhysRevD.89.064034}{\emph{Phys. Rev. D}
  {\bfseries 89} (2014) 064034}
  [\href{https://arxiv.org/abs/1308.2746}{{\ttfamily 1308.2746}}].

\bibitem{Belenchia:2016bvb}
A.~Belenchia, M.~Letizia, S.~Liberati and E.D.~Casola, \emph{{Higher-order
  theories of gravity: diagnosis, extraction and reformulation via non-metric
  extra degrees of freedom\textemdash{}a review}},
  \href{https://doi.org/10.1088/1361-6633/aaa4ab}{\emph{Rept. Prog. Phys.}
  {\bfseries 81} (2018) 036001}
  [\href{https://arxiv.org/abs/1612.07749}{{\ttfamily 1612.07749}}].

\bibitem{Sotiriou:2008rp}
T.P.~Sotiriou and V.~Faraoni, \emph{{f(R) Theories Of Gravity}},
  \href{https://doi.org/10.1103/RevModPhys.82.451}{\emph{Rev. Mod. Phys.}
  {\bfseries 82} (2010) 451} [\href{https://arxiv.org/abs/0805.1726}{{\ttfamily
  0805.1726}}].

\bibitem{DeFelice:2010aj}
A.~De~Felice and S.~Tsujikawa, \emph{{f(R) theories}},
  \href{https://doi.org/10.12942/lrr-2010-3}{\emph{Living Rev. Rel.} {\bfseries
  13} (2010) 3} [\href{https://arxiv.org/abs/1002.4928}{{\ttfamily
  1002.4928}}].

\bibitem{Jacobson:1995ab}
T.~Jacobson, \emph{{Thermodynamics of space-time: The Einstein equation of
  state}}, \href{https://doi.org/10.1103/PhysRevLett.75.1260}{\emph{Phys. Rev.
  Lett.} {\bfseries 75} (1995) 1260}
  [\href{https://arxiv.org/abs/gr-qc/9504004}{{\ttfamily gr-qc/9504004}}].

\bibitem{Eling:2006aw}
C.~Eling, R.~Guedens and T.~Jacobson, \emph{{Non-equilibrium thermodynamics of
  spacetime}}, \href{https://doi.org/10.1103/PhysRevLett.96.121301}{\emph{Phys.
  Rev. Lett.} {\bfseries 96} (2006) 121301}
  [\href{https://arxiv.org/abs/gr-qc/0602001}{{\ttfamily gr-qc/0602001}}].

\bibitem{Chirco:2009dc}
G.~Chirco and S.~Liberati, \emph{{Non-equilibrium Thermodynamics of Spacetime:
  The Role of Gravitational Dissipation}},
  \href{https://doi.org/10.1103/PhysRevD.81.024016}{\emph{Phys. Rev. D}
  {\bfseries 81} (2010) 024016}
  [\href{https://arxiv.org/abs/0909.4194}{{\ttfamily 0909.4194}}].

\bibitem{Vassilevich:2003xt}
D.V.~Vassilevich, \emph{{Heat kernel expansion: User's manual}},
  \href{https://doi.org/10.1016/j.physrep.2003.09.002}{\emph{Phys. Rept.}
  {\bfseries 388} (2003) 279}
  [\href{https://arxiv.org/abs/hep-th/0306138}{{\ttfamily hep-th/0306138}}].

\bibitem{Melmed_1988}
J.~Melmed, \emph{Conformal invariance and the regularised one-loop effective
  action}, .

\bibitem{Kennedy:1979ar}
G.~Kennedy, R.~Critchley and J.S.~Dowker, \emph{{Finite Temperature Field
  Theory with Boundaries: Stress Tensor and Surface Action Renormalization}},
  \href{https://doi.org/10.1016/0003-4916(80)90138-4}{\emph{Annals Phys.}
  {\bfseries 125} (1980) 346}.

\bibitem{tHooft:1974toh}
G.~'t~Hooft and M.J.G.~Veltman, \emph{{One loop divergencies in the theory of
  gravitation}}, {\emph{Ann. Inst. H. Poincare Phys. Theor. A} {\bfseries 20}
  (1974) 69}.

\bibitem{Wald:1984rg}
R.M.~Wald, \emph{{General Relativity}}, Chicago Univ. Pr., Chicago, USA (1984),
  \href{https://doi.org/10.7208/chicago/9780226870373.001.0001}{10.7208/chicago/9780226870373.001.0001}.

\bibitem{Oliveri:2019gvm}
R.~Oliveri and S.~Speziale, \emph{{Boundary effects in General Relativity with
  tetrad variables}},
  \href{https://doi.org/10.1007/s10714-020-02733-8}{\emph{Gen. Rel. Grav.}
  {\bfseries 52} (2020) 83} [\href{https://arxiv.org/abs/1912.01016}{{\ttfamily
  1912.01016}}].

\bibitem{Padmanabhan:2013xyr}
T.~Padmanabhan and D.~Kothawala, \emph{{Lanczos-Lovelock models of gravity}},
  \href{https://doi.org/10.1016/j.physrep.2013.05.007}{\emph{Phys. Rept.}
  {\bfseries 531} (2013) 115}
  [\href{https://arxiv.org/abs/1302.2151}{{\ttfamily 1302.2151}}].

\bibitem{Stelle:1977ry}
K.S.~Stelle, \emph{{Classical Gravity with Higher Derivatives}},
  \href{https://doi.org/10.1007/BF00760427}{\emph{Gen. Rel. Grav.} {\bfseries
  9} (1978) 353}.

\bibitem{Deruelle:2009zk}
N.~Deruelle, M.~Sasaki, Y.~Sendouda and D.~Yamauchi, \emph{{Hamiltonian
  formulation of f(Riemann) theories of gravity}},
  \href{https://doi.org/10.1143/PTP.123.169}{\emph{Prog. Theor. Phys.}
  {\bfseries 123} (2010) 169}
  [\href{https://arxiv.org/abs/0908.0679}{{\ttfamily 0908.0679}}].

\bibitem{Salvio:2018crh}
A.~Salvio, \emph{{Quadratic Gravity}},
  \href{https://doi.org/10.3389/fphy.2018.00077}{\emph{Front. in Phys.}
  {\bfseries 6} (2018) 77} [\href{https://arxiv.org/abs/1804.09944}{{\ttfamily
  1804.09944}}].

\bibitem{Donoghue:2021cza}
J.F.~Donoghue and G.~Menezes, \emph{{On quadratic gravity}},
  \href{https://doi.org/10.1393/ncc/i2022-22026-7}{\emph{Nuovo Cim. C}
  {\bfseries 45} (2022) 26} [\href{https://arxiv.org/abs/2112.01974}{{\ttfamily
  2112.01974}}].

\bibitem{Teimouri:2016ulk}
A.~Teimouri, S.~Talaganis, J.~Edholm and A.~Mazumdar, \emph{{Generalised
  Boundary Terms for Higher Derivative Theories of Gravity}},
  \href{https://doi.org/10.1007/JHEP08(2016)144}{\emph{JHEP} {\bfseries 08}
  (2016) 144} [\href{https://arxiv.org/abs/1606.01911}{{\ttfamily
  1606.01911}}].

\bibitem{Faddeev:1988qp}
L.D.~Faddeev and R.~Jackiw, \emph{{Hamiltonian Reduction of Unconstrained and
  Constrained Systems}},
  \href{https://doi.org/10.1103/PhysRevLett.60.1692}{\emph{Phys. Rev. Lett.}
  {\bfseries 60} (1988) 1692}.

\bibitem{SingerThorpe+1970+355+366}
I.M.~Singer and J.A.~Thorpe, \emph{The curvature of 4-dimensional einstein
  spaces},  in \emph{Global Analysis}, (Princeton), pp.~355--366, Princeton
  University Press (1970),
  \href{https://doi.org/doi:10.1515/9781400871230-021}{DOI}.

\bibitem{Sendouda:2011hq}
Y.~Sendouda, N.~Deruelle, M.~Sasaki and D.~Yamauchi, \emph{{Higher curvature
  theories of gravity in the ADM canonical formalism}},
  \href{https://doi.org/10.1142/S2010194511000432}{\emph{Int. J. Mod. Phys.
  Conf. Ser.} {\bfseries 01} (2011) 297}.

\bibitem{Dowker:1978md}
J.S.~Dowker and G.~Kennedy, \emph{{Finite Temperature and Boundary Effects in
  Static Space-Times}},
  \href{https://doi.org/10.1088/0305-4470/11/5/020}{\emph{J. Phys. A}
  {\bfseries 11} (1978) 895}.

\bibitem{Lanczos:1938sf}
C.~Lanczos, \emph{{A Remarkable property of the Riemann-Christoffel tensor in
  four dimensions}}, \href{https://doi.org/10.2307/1968467}{\emph{Annals Math.}
  {\bfseries 39} (1938) 842}.

\bibitem{Lovelock:1971yv}
D.~Lovelock, \emph{{The Einstein tensor and its generalizations}},
  \href{https://doi.org/10.1063/1.1665613}{\emph{J. Math. Phys.} {\bfseries 12}
  (1971) 498}.

\bibitem{Myers:1987yn}
R.C.~Myers, \emph{{Higher Derivative Gravity, Surface Terms and String
  Theory}}, \href{https://doi.org/10.1103/PhysRevD.36.392}{\emph{Phys. Rev. D}
  {\bfseries 36} (1987) 392}.

\bibitem{Davis:2002gn}
S.C.~Davis, \emph{{Generalized Israel junction conditions for a Gauss-Bonnet
  brane world}}, \href{https://doi.org/10.1103/PhysRevD.67.024030}{\emph{Phys.
  Rev. D} {\bfseries 67} (2003) 024030}
  [\href{https://arxiv.org/abs/hep-th/0208205}{{\ttfamily hep-th/0208205}}].

\bibitem{Gravanis:2002wy}
E.~Gravanis and S.~Willison, \emph{{Israel conditions for the Gauss-Bonnet
  theory and the Friedmann equation on the brane universe}},
  \href{https://doi.org/10.1016/S0370-2693(03)00555-0}{\emph{Phys. Lett. B}
  {\bfseries 562} (2003) 118}
  [\href{https://arxiv.org/abs/hep-th/0209076}{{\ttfamily hep-th/0209076}}].

\bibitem{Deruelle:2003ps}
N.~Deruelle, J.~Katz and S.~Ogushi, \emph{{Conserved charges in Einstein
  Gauss-Bonnet theory}},
  \href{https://doi.org/10.1088/0264-9381/21/8/004}{\emph{Class. Quant. Grav.}
  {\bfseries 21} (2004) 1971}
  [\href{https://arxiv.org/abs/gr-qc/0310098}{{\ttfamily gr-qc/0310098}}].

\bibitem{Moss:1996ip}
I.G.~Moss and P.J.~Silva, \emph{{BRST invariant boundary conditions for gauge
  theories}}, \href{https://doi.org/10.1103/PhysRevD.55.1072}{\emph{Phys. Rev.
  D} {\bfseries 55} (1997) 1072}
  [\href{https://arxiv.org/abs/gr-qc/9610023}{{\ttfamily gr-qc/9610023}}].

\bibitem{Luckock:1990xr}
H.~Luckock, \emph{{Mixed boundary conditions in quantum field theory}},
  \href{https://doi.org/10.1063/1.529238}{\emph{J. Math. Phys.} {\bfseries 32}
  (1991) 1755}.

\bibitem{Esposito:1997wt}
G.~Esposito, A.Y.~Kamenshchik and G.~Pollifrone, \emph{{Euclidean quantum
  gravity on manifolds with boundary}}, vol.~85, Springer, Dordrecht, Germany
  (1997),
  \href{https://doi.org/10.1007/978-94-011-5806-0}{10.1007/978-94-011-5806-0}.

\bibitem{Avramidi:1997sh}
I.G.~Avramidi and G.~Esposito, \emph{{Lack of strong ellipticity in Euclidean
  quantum gravity}},
  \href{https://doi.org/10.1088/0264-9381/15/5/006}{\emph{Class. Quant. Grav.}
  {\bfseries 15} (1998) 1141}
  [\href{https://arxiv.org/abs/hep-th/9708163}{{\ttfamily hep-th/9708163}}].

\bibitem{Moss:2013vh}
I.G.~Moss, \emph{{BRST-invariant boundary conditions and strong ellipticity}},
  \href{https://doi.org/10.1103/PhysRevD.88.104039}{\emph{Phys. Rev. D}
  {\bfseries 88} (2013) 104039}
  [\href{https://arxiv.org/abs/1301.0717}{{\ttfamily 1301.0717}}].

\bibitem{Fursaev:2015wpa}
D.~Fursaev, \emph{{Conformal anomalies of CFT\textquoteright{}s with
  boundaries}}, \href{https://doi.org/10.1007/JHEP12(2015)112}{\emph{JHEP}
  {\bfseries 12} (2015) 112}
  [\href{https://arxiv.org/abs/1510.01427}{{\ttfamily 1510.01427}}].

\bibitem{Sonego:1993fw}
S.~Sonego and V.~Faraoni, \emph{{Coupling to the curvature for a scalar field
  from the equivalence principle}},
  \href{https://doi.org/10.1088/0264-9381/10/6/015}{\emph{Class. Quant. Grav.}
  {\bfseries 10} (1993) 1185}.

\bibitem{DeSimone:2008ei}
A.~De~Simone, M.P.~Hertzberg and F.~Wilczek, \emph{{Running Inflation in the
  Standard Model}},
  \href{https://doi.org/10.1016/j.physletb.2009.05.054}{\emph{Phys. Lett. B}
  {\bfseries 678} (2009) 1} [\href{https://arxiv.org/abs/0812.4946}{{\ttfamily
  0812.4946}}].

\bibitem{PhysRevD.25.1019}
B.L.~Nelson and P.~Panangaden, \emph{Scaling behavior of interacting quantum
  fields in curved spacetime},
  \href{https://doi.org/10.1103/PhysRevD.25.1019}{\emph{Phys. Rev. D}
  {\bfseries 25} (1982) 1019}.

\bibitem{Gustafson1998DomainD}
K.E.~Gustafson, \emph{Domain decomposition, operator trigonometry, robin
  condition},  1998.

\end{thebibliography}\endgroup
\bibliographystyle{JHEP}

\end{document}